\documentclass[superscriptaddress,twocolumn,showpacs]{revtex4}
\usepackage{graphicx}
\usepackage{color}
\usepackage{dcolumn}
\usepackage{amsmath}
\usepackage{dsfont}

\newcommand{\be}{\begin{equation}}
\newcommand{\ee}{\end{equation}}

\begin{document}

\title{Nonlinear multi-core waveguiding structures with balanced gain and loss}

\author{Alejandro J. Mart\'inez}

\affiliation{Departamento de F\'isica, MSI-Nucleus on Advanced Optics,
and Center for Optics and Photonics (CEFOP), Facultad de Ciencias,
Universidad de Chile, Santiago, Chile}

\affiliation{Oxford Center for Industrial and Applied Mathematics,
Mathematical Institute, University of Oxford, Oxford OX2 6GG, United Kingdom}

\author{Mario I. Molina}

\affiliation{Departamento de F\'isica, MSI-Nucleus on Advanced Optics,
and Center for Optics and Photonics (CEFOP), Facultad de Ciencias,
Universidad de Chile, Santiago, Chile}

\author{Sergei K. Turitsyn}

\affiliation{Aston Institute of Photonic Technologies,
Aston University, Birmingham B4 7ET, United Kingdom}

\author{Yuri S. Kivshar}

\affiliation{Nonlinear Physics Center, Research School of Physics and
Engineering, Australian National University, Canberra ACT 0200,
Australia}


\begin{abstract}
We study existence, stability, and dynamics of linear and nonlinear stationary modes propagating
in radially symmetric multi-core waveguides with balanced gain and loss. We demonstrate that, in general,
the system can be reduced to an effective ${\cal PT}$-symmetric dimer with asymmetric coupling.
In the linear case,  we find that there exist two modes with real propagation constants before
an onset of the ${\cal PT}$-symmetry breaking while other modes have always the propagation constants
with nonzero imaginary parts. This leads to a stable (unstable) propagation of the modes when gain
is localized in the core (ring) of the waveguiding structure.  In the case of nonlinear response,  we
show that an interplay between nonlinearity, gain, and loss induces a high degree of instability,
with only small windows in the parameter space where quasi-stable propagation is observed. We propose
a novel stabilization mechanism based on a periodic modulation of both gain and loss along the propagation
direction that allows bounded light propagation in the multi-core waveguiding structures.
\end{abstract}

\maketitle

\section{INTRODUCTION}

During last decade, many efforts have been devoted to the study of photonic structures consisting of coupled waveguides with gain and loss~\cite{RuschhauptL171,Ganainy2632} which offer interesting novel possibilities for shaping optical beams in comparison with traditional conservative or low-loss structures. Many of such structures can be constructed as optical analogues of the complex space-time potentials possessing the so-called parity-time ($\mathcal{PT}$)-symmetry, which can have an entirely real eigenvalue spectrum, meaning the energy conservation of optical eigenmodes. The symmetry here can be interpreted as an optical equivalent of the ${\cal PT}$ symmetry in quantum mechanics~\cite{bender, Bender:2003-1095:AMJP, Bender:2007-947:RPP}.

The first experimental demonstrations of the ${\cal PT}$-symmetric effects in optics were in two-waveguide directional linear couplers composed of waveguides with gain and loss~\cite{Guo:2009-93902:PRL, Ruter:2010-192:NPHYS}. Theoretical analysis suggests that  such couplers, operating in the nonlinear regime, can be used for the all-optical signal control~\cite{Ramezani:2010-43803:PRA,Coupler3,ptdimer}. Arrays of the ${\cal PT}$-symmetric couplers were proposed as a feasible means of control of the spatial beam dynamics, including the formation and switching of spatial solitons~\cite{Dmitriev:2010-2976:OL, Zheng:2010-10103:PRA, Suchkov:2011-46609:PRE}.

Recently, a new theory of coherent propagation and power transfer in low-dimension array of coupled nonlinear waveguides has been suggested by Turitsyn {\em et al.}~\cite{turit,turit2}, where it was demonstrated that in the array with the central core stable steady-state coherent multi-core propagation is possible only in the nonlinear regime, with a power-controlled phase matching. This finding opens novel opportunities to explore multi-core waveguiding systems, however it also puts a question about the stability of such waveguiding structures in the presence of gain and loss. We notice that, apart from being an interesting physical system, a multi-core optical fibre is now actively studied in the context of the spatial division multiplexing, the technology of transmitting information over separate spatial channels. The spatial division multiplexing enables the up-scaling of the  capacity per-fibre that is a critical challenge in the modern optical communications \cite{MCF1,MCF2}. Multi-core optical fibres are also studied in the field of powerful fibre lasers \cite{MCF3}, where gain is an important feature of the system. Multi-core waveguiding systems may be useful when nonlinear effects limit the power that can be transmitted in a single waveguide. In this case, multi-core waveguiding system can operate in the regime when the light power in each core is below the level of non-desirable physical effects, while coherence is provided by the coupling between the waveguide cores, allowing for the coherent combining of the total power after delivering the signal to destination. Gain and loss are both important in such multi-core optical fibre systems.

In this paper, we study both linear and nonlinear dynamics in multi-core waveguiding systems suggested earlier
in Refs.~\cite{turit,turit2}, but in the presence of balanced gain and loss, when the system operates as a multi-core optical coupler (see Fig.~1). First, we analyze all regime when the system can be transformed into the ${\cal PT}$-symmetric multi-core couplers and study its stability. Then we suggest how to achieve the bounded
dynamics in the nonlinear regime by modulating both gain and loss.

\begin{figure}
\includegraphics[height=4.cm]{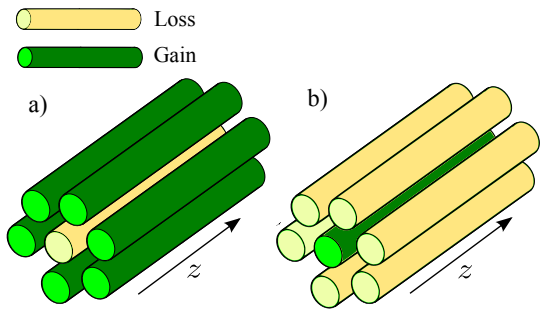}
\caption{Schematic of the waveguiding structure of a radially symmetric
multi-core waveguide array with balanced gain and loss.}
\label{fig1}
\end{figure}

The paper is organized as follows. In Secs. II and III we describe a general model and focus on the study
of the linear regime finding the critical parameters for gain and loss when the ${\cal PT}$ symmetry breaks.
In Sec.~IV  we discuss the reduction of the multi-core coupler to an asymmetric waveguide dimer in the presence
of nonlinearity. We find all nonlinear modes and analyze their stability. In Sec.~V we study the dynamical
evolutions of the modes and discuss their numerical stability, culminating with a proposal
to stabilize the system by means of spatially-periodic gain and loss. Section VI concludes the paper.

\section{Model}

We consider a multi-core waveguide array composed of $N$ identical waveguides arranged in a circular geometry, as shown in Fig.~1. We assume that all waveguides are identical, and they are characterized by the propagation constant $\epsilon_{1}$, with gain/loss parameter $\rho_{1}$. In addition, we include  a central waveguide with the propagation constant $\epsilon_{0}$, and gain/loss parameter $\rho_{0}$. The nonlinear parameter $\gamma$ also can be different for the central and peripheral cores, but we assume it to be the same in the presented analysis.  In the coupled-modes formalism applied here, we assume the interaction of  the nearest neighbors for the waveguides on the ring,  and write the evolution equations for the mode amplitudes in the form,
\begin{eqnarray}
-i\frac{dA}{dz}&=&(\epsilon_{0} + i \rho_{0}) A+C_{0} \sum_{j=1}^{N}
B_{j} + \gamma |A|^{2}A,\label{eq:1}\\
-i\frac{dB_{j}}{dz}&=& (\epsilon_{1} + i \rho_{1}) B_{j}
+C_{1} (B_{j+1}+B_{j-1})\label{eq:2}\\&&+ C_{0}A
+ \gamma |B_{j}|^{2}B_{j},\nonumber
\end{eqnarray}
where $A$ is the amplitude of the electric field in the core waveguide, $B_{j}$ is the amplitude in
the $j$-th waveguide on the ring, with the conditions $B_0=B_N$ and $B_{N+1}=B_1$, $\gamma$
being the Kerr nonlinearity coefficient, and $C_{0,1}$ being the coupling coefficients
of the modes of different waveguides.

The coupling coefficients $C_{0}$ and $C_{1}$ are not independent. For a circular array of $N$ waveguides, the distance between the nearest-neighbor waveguides in the ring $L$, and the distance from the center core to the ring $R_{0}$, are related by the condition $L = 2 R_{0} \sin(\pi/N)$. Using the fact that $C_{0}\sim \exp(-\mu R_{0})$ and $C_{1}\sim
\exp(-\mu L)$, where $\mu$ depends on physical parameters such as, geometry of waveguides or their refractive indices,
we obtain
\be
\frac{C_{1}}{C_{0}}=\exp\left\{\mu R_{0} \left[1-2
\sin(\pi/N)\right]\right\}.
\label{C1overC0}
\ee
Thus for $N<6$, we have $C_{1}<C_{0}$, while for $N \geq 6$, we obtain $C_{1}\geq
C_{0}$, where the equality is satisfied only when $N=6$. We notice that this corresponds to the recently developed 7-core multi-core fibre  actively studied in optical communication \cite{MCF1,MCF2}, where typical examples of the parameters can be found.

\section{Linear regime}

\subsection{Eigenvalues and linear modes}

\begin{figure}
\includegraphics[height=5.cm]{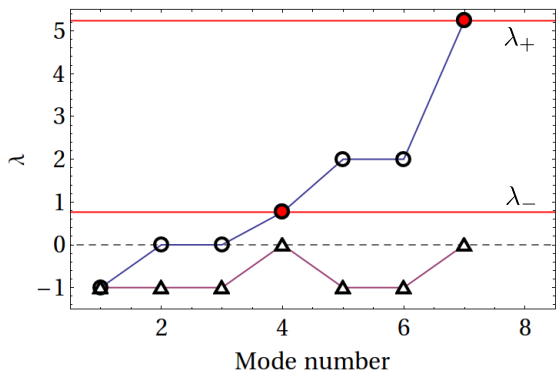}
\caption{(Color online) Eigenvalues of Eqs. (1),(2) for $N=6$, $\gamma=0$ and $\rho = 1$.
Circles (triangles) show the real (imaginary) part of the eigenvalues.
Red lines show the pure real eigenvalues associated with the ${\cal PT}$-symmetry in the reduced
system.}
\label{fig2}
\end{figure}

First, we consider the linear case when $\gamma=0$. According to Ref.~\cite{efre}, in this case the system described
by Eqs. (\ref{eq:1})-(\ref{eq:2}) has only two rotational invariant
modes, such that $B_{n}=B$ for all $n$, and these modes are associated
with pure real eigenvalues for
$|\rho|<\rho_{c}$:
\be
\lambda^{\pm} = \epsilon_{1} + 2 C-{1} \pm \sqrt{N C_{0}^{2}-
\rho^{2}},\label{eq:4}
\ee
where $\rho_{0}=\rho=-\rho_{1}$ and $\rho_{c}=\sqrt{N} C_{0}$. The other modes
correspond to waves without field in the central guide. As a
consequence, their
eigenvalues are exactly the eigenvalues of a ring \cite{efre}, i.e., a
one-dimensional chain with periodic boundary condition:
\be
\lambda_{\nu} = -i\rho + 2C_1\cos\left(\frac{2\pi \nu}{N}\right),\; \nu = 0,
1,{\ldots}, N-2.\label{eq:5}
\ee
Moreover, they are organized in pair of degenerated modes.
Figure~\ref{fig2} shows examples of the linear eigenvalues for the case $N=6$.
Note that these $N-1$ eigenvalues have
imaginary part
equal to $-\rho$, which means that the linear modes associated with
them can be written as
$B_{n,\nu}(z) = b_n
e^{2iC_1\cos\left(\frac{2\pi \nu}{N}\right) z}e^{\rho z}$, where $b_n$
is the profile of the mode. The dominant behavior is given by the real
exponential term $e^{\rho z}$, such that the optical field $B_{n,\nu}$
either goes to zero for $\rho < 0$ or is
unbounded for $\rho>0$, as $z\rightarrow \infty$.

\begin{figure}
\includegraphics[height=2.3cm]{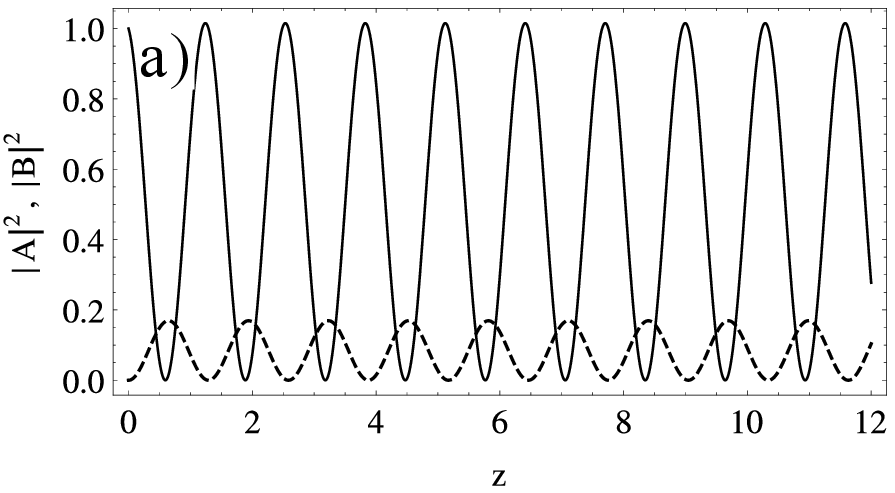}
\includegraphics[height=2.3cm]{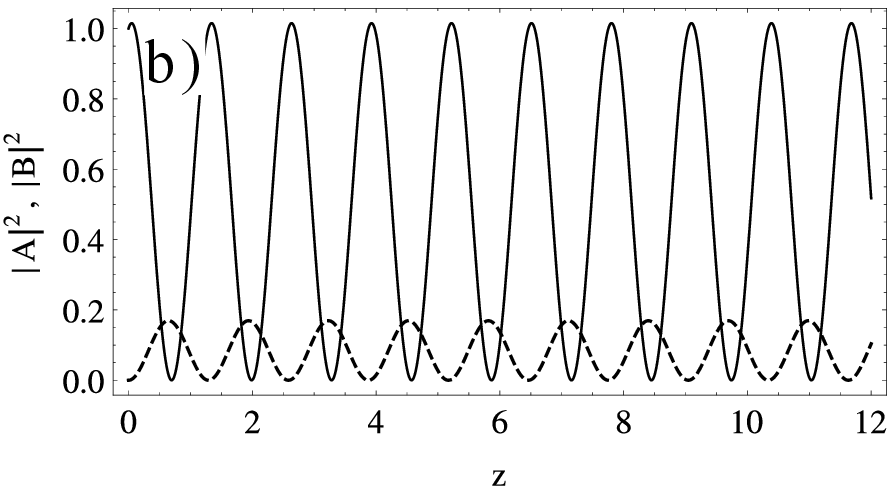}\\
\includegraphics[height=2.3cm]{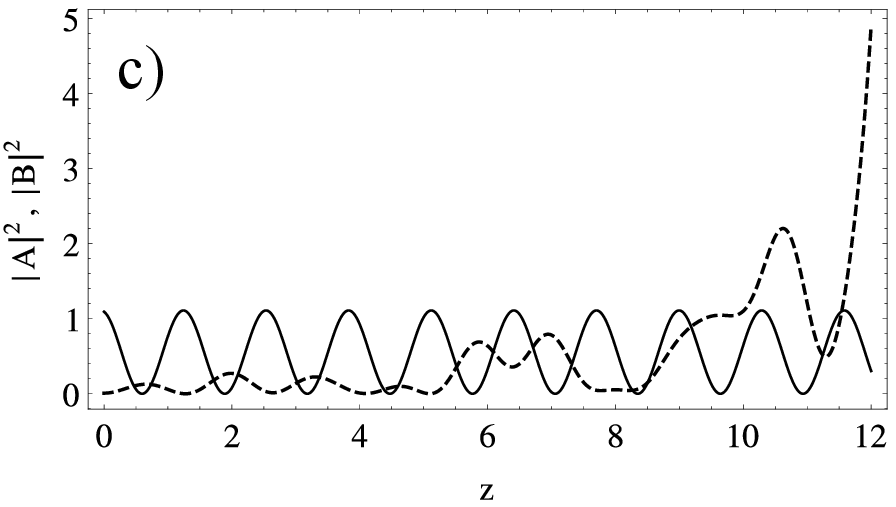}
\includegraphics[height=2.3cm]{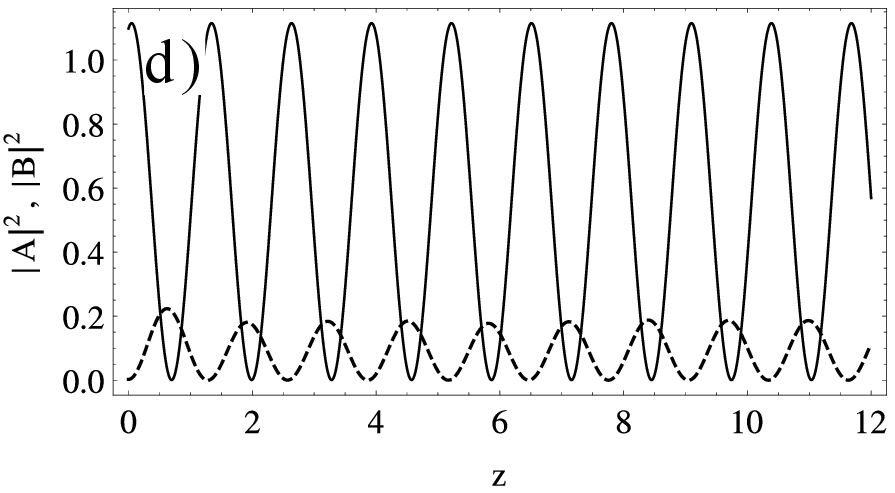}\\
\includegraphics[height=2.3cm]{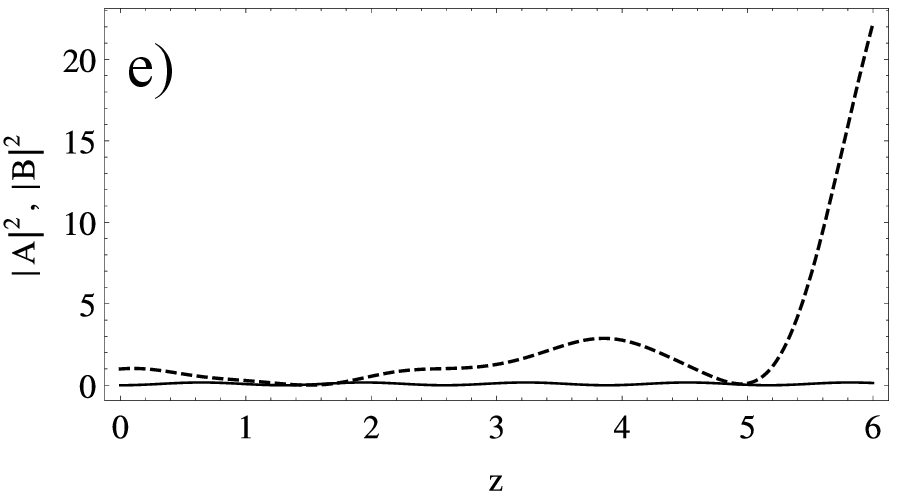}
\includegraphics[height=2.3cm]{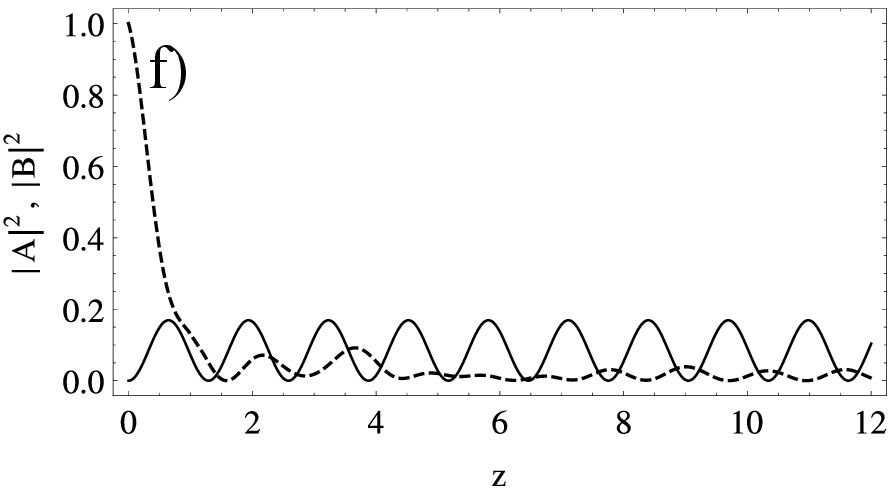}
\caption{Examples of numerical integration of Eqs. (1),(2)
in the linear regime ($\gamma = 0$) for $N=6$ and $|\rho|<\rho_c$.
Continuous and dashed lines are associated with $A(z)$ and $B_1(z)$,
respectively. Left and right columns correspond to $\rho>0$ and $\rho<0$ cases,
respectively. The initial conditions are: (a)-(d) $A(0)=1$ and $B_j(0)=0$, and (e)-(f) $A(0)=B_{j>1}(0)=0$ and $B_{1}(0)=1$. Furthermore, in cases (c) and (d) a small white noise has been added to the initial condition.}
\label{fig3}
\end{figure}

From a dynamical point of view, any initial condition of the general
system of the form $A(0) = a_0$ and $B_n(0)=b_0$ excites only the modes
related to equation (\ref{eq:4}). On the other hand, for a completely
arbitrary initial
condition, namely, $A(0)=a_0$ and $B_n(0)=b_n$, we have to consider the
contribution of each mode in the form
\begin{eqnarray}
\{A(z),B_n(z)\} &=& \{\alpha^+,\beta_n^+\}e^{i\lambda^+ z}+
\{\alpha^-,\beta_n^-\}e^{i\lambda^- z}\nonumber\\&&+
\sum_{\nu=0}^{N-2}\{\alpha^{\nu},\beta_n^{\nu}\}e^{i\lambda_{\nu} z},\label{eq:6}
\end{eqnarray}
where the coefficients $\alpha^+,\beta^+,{\ldots}$ correspond to the
projection of the initial condition over the appropriate eigenvector.
Equation (\ref{eq:4}) and (\ref{eq:5}) implies that the dynamic, given
by (\ref{eq:6}), remains bounded only
for $-\sqrt{N}C_0<\rho<0$, approaching asymptotically a situation where only
the modes associated with $\lambda^{\pm}$ have a significant role.
Fig.~\ref{fig3} shows some dynamical evolutions for different initial
conditions in each regimen of $\rho$. Also, we include some numerical
simulations with a noisy (white noise) initial condition in order to
have a finite contribution of every linear mode to the dynamics.

\subsection{Reduction to an effective waveguide dimer}

Since the effective dynamics of the system quickly converges to
that of a dimer for $\rho<0$, let us simplify the problem and
work with a nondegenerate dimer from the outset~\cite{turit}:
\begin{eqnarray}
-i\frac{dA}{dz}&=& (\epsilon_{0} + i \rho_{0}) A+ N C_{0} B,\\
-i\frac{dB}{dz}&=& (\epsilon_{1} + i \rho_{1}) B+ C_{0} A + 2
C_{1} B.
\label{2}
\end{eqnarray}

Note that, solutions described by this reduction are invariant under
discrete rotations in $2\pi n/N$ ($n\in \mathds{Z}$) respect to the
central waveguide.

We pose a solution
of the form $A(z)=a \exp(i \lambda z), B(z)=b \exp(i \lambda z)$. This
leads to:
\begin{eqnarray}
(-\lambda+\epsilon_{0} + i \rho_{0})a+N C_{0} b &=&0,\label{eq:9}\\
C_{0} a + (-\lambda+\epsilon_{1}+2 C_{1}+i \rho_{1}) b &=&0.\label{eq:10}
\end{eqnarray}
Examination of the determinant of the system reveals that in order to
have $\lambda$ real, one needs to impose
$\rho_{0}=-\rho_{1}=\rho$ and $\epsilon_{0}-\epsilon_{1}=2 C_{1}$, leading
to the  propagation constants:
\be
\lambda^{\pm} = \epsilon_{1}+2 C_{1} \pm \sqrt{N C_{0}^{2} - \rho^{2}}.
\label{eq:11}
\ee
Thus, the critical gain and loss  parameter value is $\rho_{c}^2=N C_{0}^2$.

The eigenvectors are given by
\be
\left\{a^{\pm},b\right\} = \left\{ \frac{i\rho\pm \sqrt{N
C_{0}^{2}-\rho^{2}}}{C_{0}}, 1\right\},\label{eq:12}
\ee
and satisfy $|a^{\pm}|^{2} = N |b|^{2}$ when $|\rho|\leq\rho_c$ and,
$C_0^2|a^{\pm}|^2 =
\left(\rho\pm \sqrt{\rho^2-NC_0^2}\right)^2|b|^2$
when $|\rho|>\rho_c$.

\begin{figure}
\includegraphics[height=8.4cm]{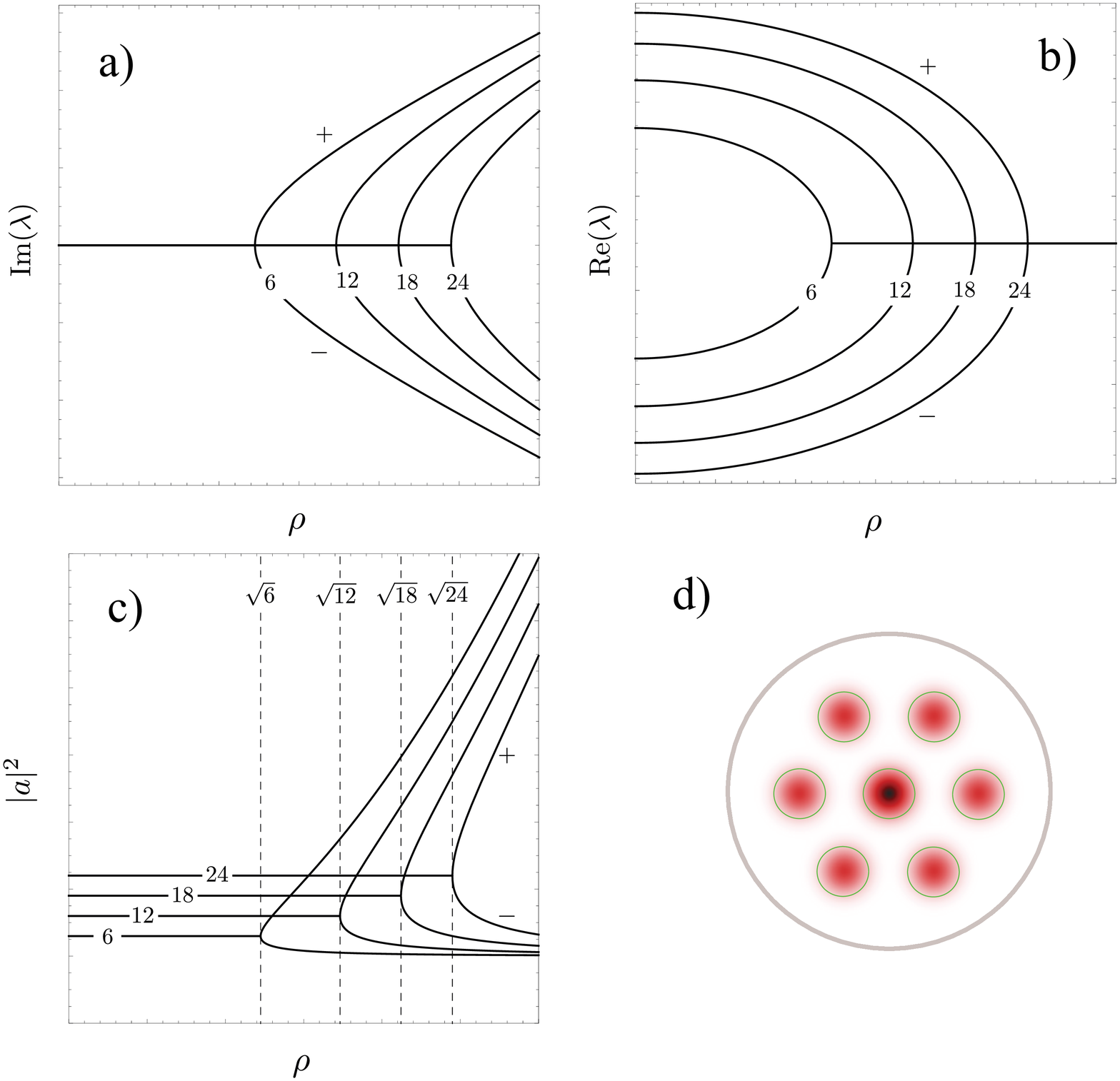}
\caption{(Color online) a) and b) show the imaginary and real part of the eigenvalues
described in Eqs.(\ref{eq:9})-(\ref{eq:10}), respectively. c) shows the intensity of the
optical field in the central waveguide, which is given
by Eq.(\ref{eq:11}). d)
shows an example of the intensity distribution of the linear modes in
a system with $N=6$. a), b) and c) were calculated for $N=6, 12, 18, 24$.}
\label{hola}
\end{figure}

At least in the dimer reduction, the requirement over the system
parameters leads to a ${\cal PT}$-symmetric dimer, such as in Refs.~\cite{ptdimer,
ptdimer2}. Thus, even though the Hamiltonian is non-Hermitian, its eigenvalues
will be in $\mathds{R}$ until the onset of ${\cal PT}$-symmetry-breaking~\cite{bender}.
For any other solution that could not be described by this reduction,
the system intrinsically does not satisfy ${\cal PT}$-symmetry.
Fig.~\ref{hola} shows the propagation constants as function of $\rho$, the
bifurcation diagram $|a(\rho)|^2$, and an example of the intensity
distribution of the linear modes.

\subsection{System dynamics}

We consider the linear dynamics of an arbitrary initial condition:
$A(0)=a_0, B(0)=b_0$. One expands
\be
A(z) = \alpha^+ \exp(i \lambda^{+} z) + \alpha^- \exp(i \lambda^{-} z).
\ee

In general, the periodicity of $A(z)$ along $z$ depends on the ratio
between $\lambda^+$ and $\lambda^-$. For instance, if
$\lambda^+/\lambda^- \in \mathds{Z}$ then $A(z)$ will be
periodic. Otherwise, it will be aperiodic. Nonetheless,
the intensity of the field is periodic and given by
\be
|A(z)|^2 =
|\alpha^+|^2+|\alpha^-|^2+\kappa\sin\left((\lambda^+-\lambda^-)z
+ \phi\right),
\label{average}
\ee
where $\kappa$ and $\phi$ are the amplitude and phase of
the periodic oscillation around the average value
$\left<|A(z)|^2\right>_z = |\alpha^+|^2+|\alpha^-|^2$ of the
intensity, respectively. Both quantities
are functions of the initial condition $\{a_0,b_0\}$. Equation
(\ref{average}) means that the intensity is periodic for all
$\lambda^+\neq \lambda^-$. The characteristic
propagation constant is given by $\lambda_c =
\lambda^+-\lambda^- = 2\sqrt{N C_0^2-\rho^2}$.
In particular, when $\lambda^+ = \lambda^-$ ($\rho
= \rho_c$) the intensity remains constant.

One interesting case corresponds to the excitation only at the core,
i.e.,  $a_0=1$ and $b_0=0$. In this case, the parameters of the
expansion are
\begin{eqnarray}
\alpha^+&=&\left( {\sqrt{N C_{0}^2-\rho^2}+i \rho\over{2 \sqrt{N
C_{0}^2-\rho^2}}}\right),\\ \alpha^-&=&\left( {\sqrt{N
C_{0}^2-\rho^2}-i \rho\over{2 \sqrt{N  C_{0}^2-\rho^2}}}\right),
\end{eqnarray}
and both quantities are singular when $\rho=rho_c$. On the other
hand, unlike the case without gain and loss, in general
there is no conservation of
power $P=|A(z)|^2 + |B(z)|^2$, and the power transfer between the
two sites is asymmetrical. This is not a physical problem, rather it is just
a mathematical consequence of the reduction. The
total power of the entire system, which is conserved in this case,
is $P_N = |A(z)|^2+N|B(z)|^2$.

\section{Nonlinear regime}

\subsection{Structure of nonlinear modes}

\begin{figure}
\includegraphics[height=5.cm]{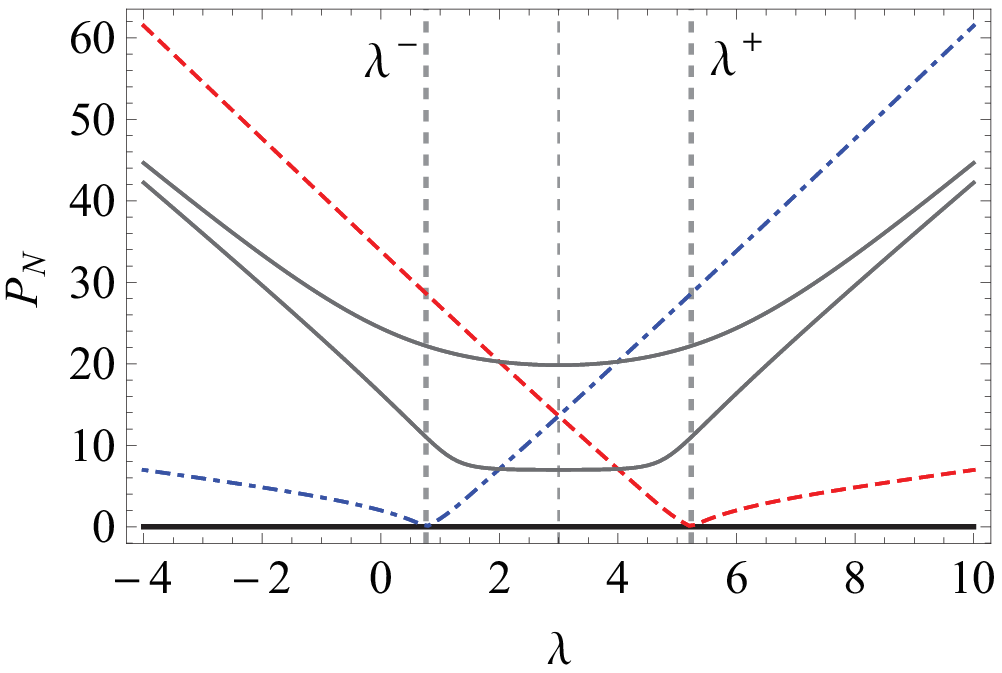}\\
\includegraphics[height=2.8cm]{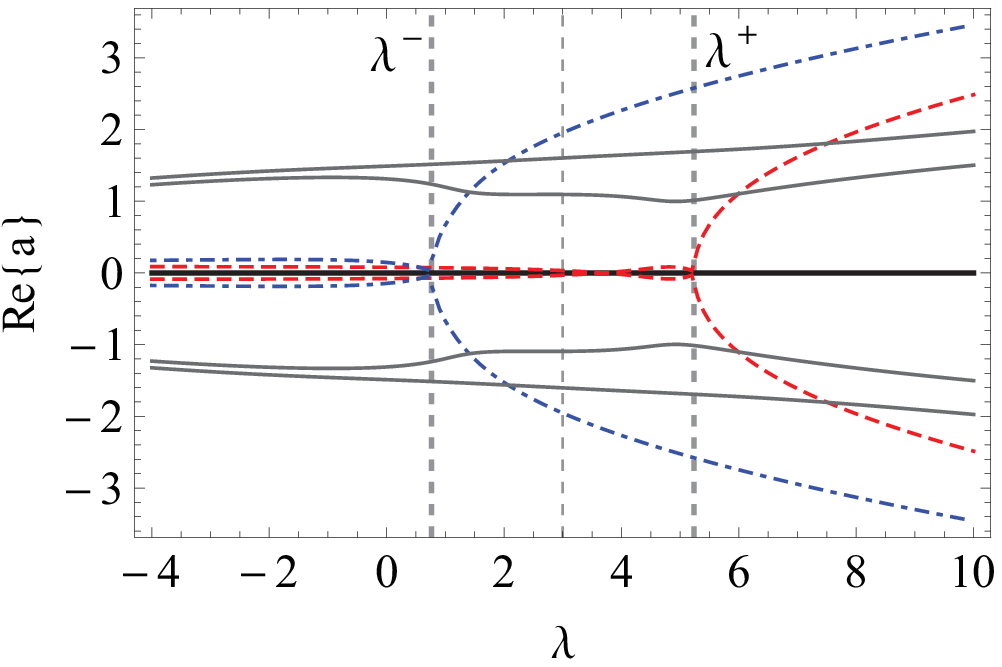}
\includegraphics[height=2.8cm]{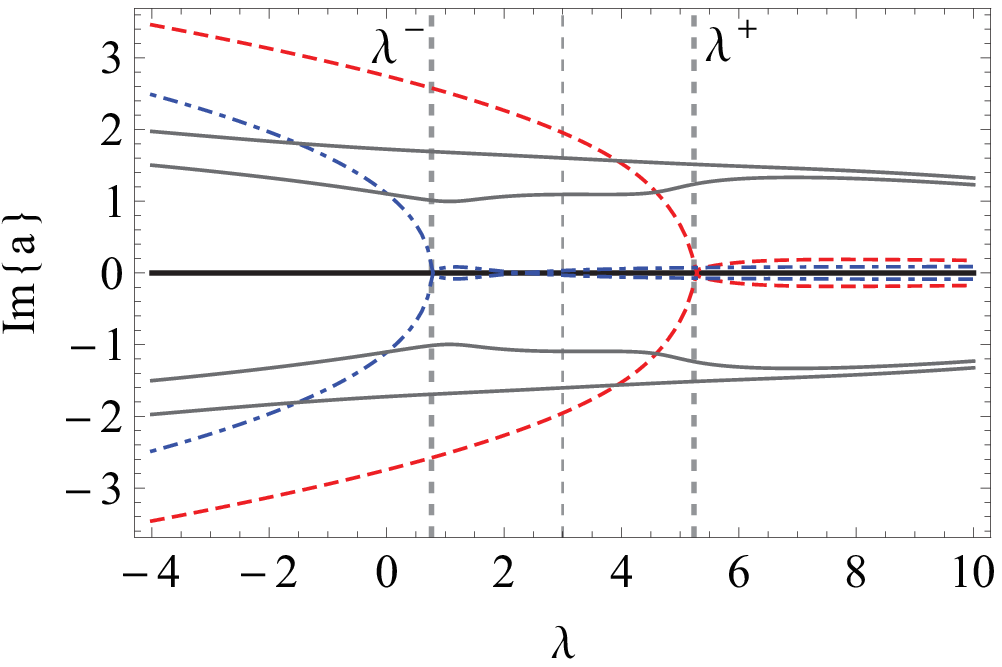}\\
\includegraphics[height=2.8cm]{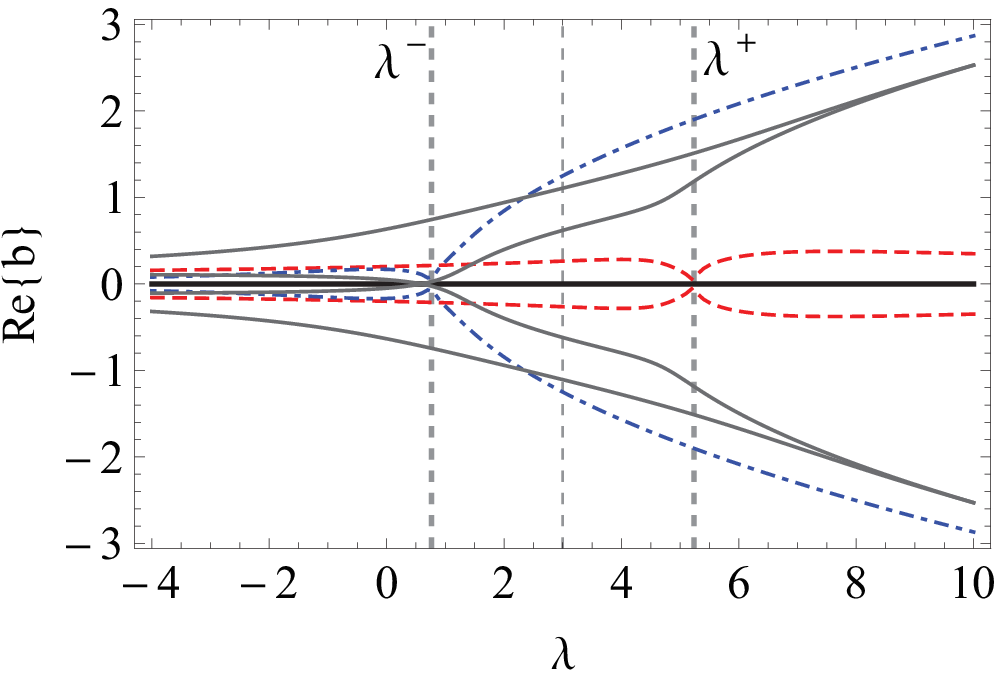}
\includegraphics[height=2.8cm]{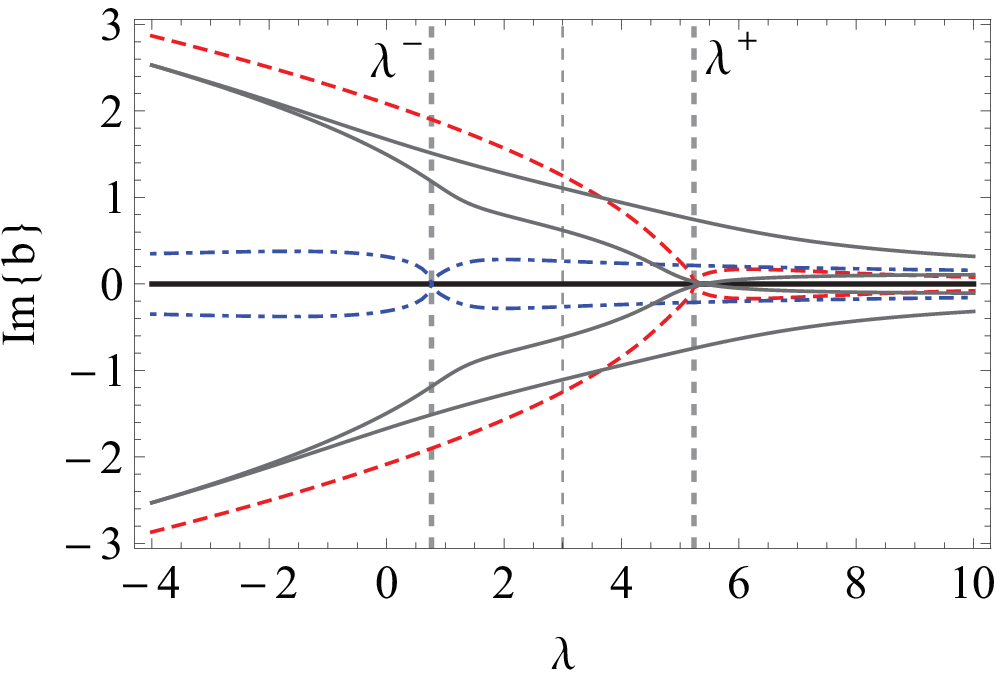}
\caption{(color online) $P_N$, $\text{Re}\{a\}$, $\text{Re}\{b\}$,
$\text{Im}\{a\}$, and $\text{Im}\{b\}$ associated with the 9 solutions
of Eqs. (\ref{eq:16}), (\ref{eq:17}) as function of the propagation constant $\lambda$.
Red and blue lines denote the modes described
by Eq. (\ref{eigenvectors}). Thick vertical lines denote
the propagation constants $\lambda^{\pm}$, while thin vertical lines are related to $(\lambda^+
+\lambda^-)/2$.}\label{hola2}
\end{figure}

Let us continue working within the dimer reduction.
In the presence of nonlinear effects, the equations read
\begin{eqnarray}
-i\frac{da}{dz}&=& (\epsilon_{0} + i \rho_{0}) a+N C_{0} b +
\gamma |a|^2 a,\label{new16}\\
-i\frac{db}{dz}&=& (\epsilon_{1} + i \rho_{1}) b+C_{0} a + 2
C_{1} b + \gamma |b|^{2} b\label{new17}.
\label{2}
\end{eqnarray}
A stationary state solution
$a(z)=a \exp(i \lambda z)$, $b(z)=b \exp(i \lambda z)$, leads to the
equations
\begin{eqnarray}
(-\lambda+\epsilon_{0}+ i \rho + \gamma |a|^2)a + N C_{0}
b&=&0,\label{eq:16}\\
C_{0} a + (-\lambda+\epsilon_{1}+2 C_{1}-i \rho+\gamma
|b|^2)b&=&0.\label{eq:17}
\end{eqnarray}
The transformation: $\{m^+,\lambda^+,\rho,\gamma\} \rightarrow
\{m^-,\lambda^-,-\rho,-\gamma\}$, leaves Eqs. (\ref{eq:16}), (\ref{eq:17})
invariant.  Thus, we will analyze only the case with
self-focusing nonlinearity ($\gamma>0$).
These above equations have exactly 9 complex solutions, one of them being the zero
(trivial) solution. The other solutions are organized in pairs with
identical relation between $P_N$ and $\lambda$.
Fig.~\ref{hola2} shows the power vs propagation constant diagram for
these modes, as well as their real and imaginary parts as functions
of the propagation constant $\lambda$.
We note that, there are four nonlinear branches that emerge
exactly from the propagation constant associated with the linear modes
described by Eq.~(\ref{eq:12}).

In order to simplify the description of these solutions,
we introduce a shift in the propagation constant
$\lambda\rightarrow \lambda+\epsilon_0 = \lambda + \left(\lambda^+
+\lambda^-\right)/2$, with $\epsilon_0=\epsilon_1+2C_1$. Thus, Eqs.
(\ref{eq:16}) and (\ref{eq:17}) read
\begin{equation}
 \lambda  \vec{v}=\left(
\begin{array}{cc}
i\rho + \gamma|a|^2&NC_0\\
C_0 & -i\rho + \gamma|b|^2
\end{array}
\right)\vec{v},\label{eq:18}
\end{equation}
where $\vec{v} = \left(a,b\right)^T$, and $T$ denotes the transpose.

Equation (\ref{eq:18}) has a {\it phase invariance}, i.e., it
is invariant under a global phase shift $\vec{v}\rightarrow
\vec{v}e^{i\theta}$. This is connected with the fact that the
conservation of the power $P_N$ is a necessary condition in order to
have stationary fields. Actually, it is easy to show that to avoid
fluctuations in the power along the propagation direction, $P_N$ must
satisfy
\begin{equation}
\frac{dP_N}{dz} = - 2\rho\left(|a|^2-N|b|^2\right) = 0,\label{eq:19}
\end{equation}
which means that solutions of \eqref{eq:16} and \eqref{eq:17}
must satisfy $|a|^2=N|b|^2$. This relation can equivalently be derived
directly from Eqs. \eqref{eq:16} and
\eqref{eq:17}  through imposing condition of real eigenvalues
$\text{Im}(\lambda)=0$. Thus, there are pairs of solutions of the form $\pm\{a,b\}$ as is
shown in Fig. \ref{hola2}.

Moreover, Eqs. \eqref{eq:16} and \eqref{eq:17}
represent a nonlinear spectral problem with solutions $a$, $b$, and $\lambda$ been
function of $N$, $\epsilon_{0,1}$, $C_{0,1}$, $\gamma$ and $\rho$.
Let us denote $\text{Re}(\lambda)=\lambda_R$, $\text{Im}(\lambda)=\lambda_I$ and
introduce $\Gamma= b/a$. Now, we look for stationary solutions assuming
that $|a|^2=N|b|^2$ is satisfied and $P_N=|a|^2 +N
|b|^2=const$, thus we get from Eqs. (\ref{eq:16}) and
(\ref{eq:17}) that $\Gamma_I=\text{Im}(\Gamma)= - \rho/(N C_0)$ and
$\Gamma_R=\text{Re}(\Gamma)=\pm \sqrt{1/N- \rho^2/(N^2 C_0^2)}$. Thereby,
the nonlinear solutions read:
\be
m^{\pm}=\{a^{\pm}, b\} = \sqrt{\frac{P_N}{2 N}} \, \left\{ {i \rho\pm \sqrt{N
C_0^2-\rho^2}\over{C_{0}}},1\right\},
\label{eigenvectors}
\ee
and the propagation constants are:
\be
\lambda_{\gamma}^{\pm} = \epsilon_{0} + \gamma \frac{P_N}{2} \pm \sqrt{N
C_{0}^2-\rho^2},
\label{lambda}
\ee
which correspond to the nonlinear continuation of the linear modes
given by Eq. \eqref{eq:12}. These naturally satisfy $|a^{\pm}|^2 = N
|b|^2$ (when $|\rho|<\rho_c$) by construction.
Furthermore, The ${\cal PT}$ symmetry-breaking critical gain and loss parameter is the same as
before, $\rho_{c}^2=N C_{0}^2$. Additionally, we note that while nonlinearity
induces a shift in the propagation constant of these modes, their
envelope remain unchanged, except by a constant factor that depends
on the total power and the number of waveguides in the multi-core array.

\subsection{Stability analysis}

To examine the linear stability of the nonlinear modes given by Eq. (\ref{eigenvectors}), we
introduce small perturbations and write the amplitudes in the form, $A(z)\rightarrow (a+\delta_{0}(z))\exp(i
\lambda z)$ and $B\rightarrow (b+\delta_{1}(z))\exp(i \lambda z)$.
After inserting this into Eqs. (\ref{new16}), (\ref{new17}), we obtain in the first order in
$\delta_{0}, \delta_{1}$ the following linear equations,
\begin{eqnarray}
-i\frac{d\delta_{0}}{dz}&=&  (-\lambda+\epsilon_{0}+ i \rho)
\delta_{0} + N C_{0}\delta_1\label{26}\\
&&+\gamma a^2 \delta_{0}^{*} + 2\gamma
|a|^2 \delta_{0},\nonumber\\
-i\frac{d\delta_{1}}{dz}&=& (-\lambda +\epsilon_{1} + 2 C_{1}-i
\rho) \delta_{1}+C_{0}\delta_{0}\label{27}\\
&&+\gamma b^2
\delta_{1}^{*}+2 \gamma |b|^2 \delta_{1}.\nonumber\label{15}
\end{eqnarray}

\begin{figure}
\includegraphics[height=3.5cm]{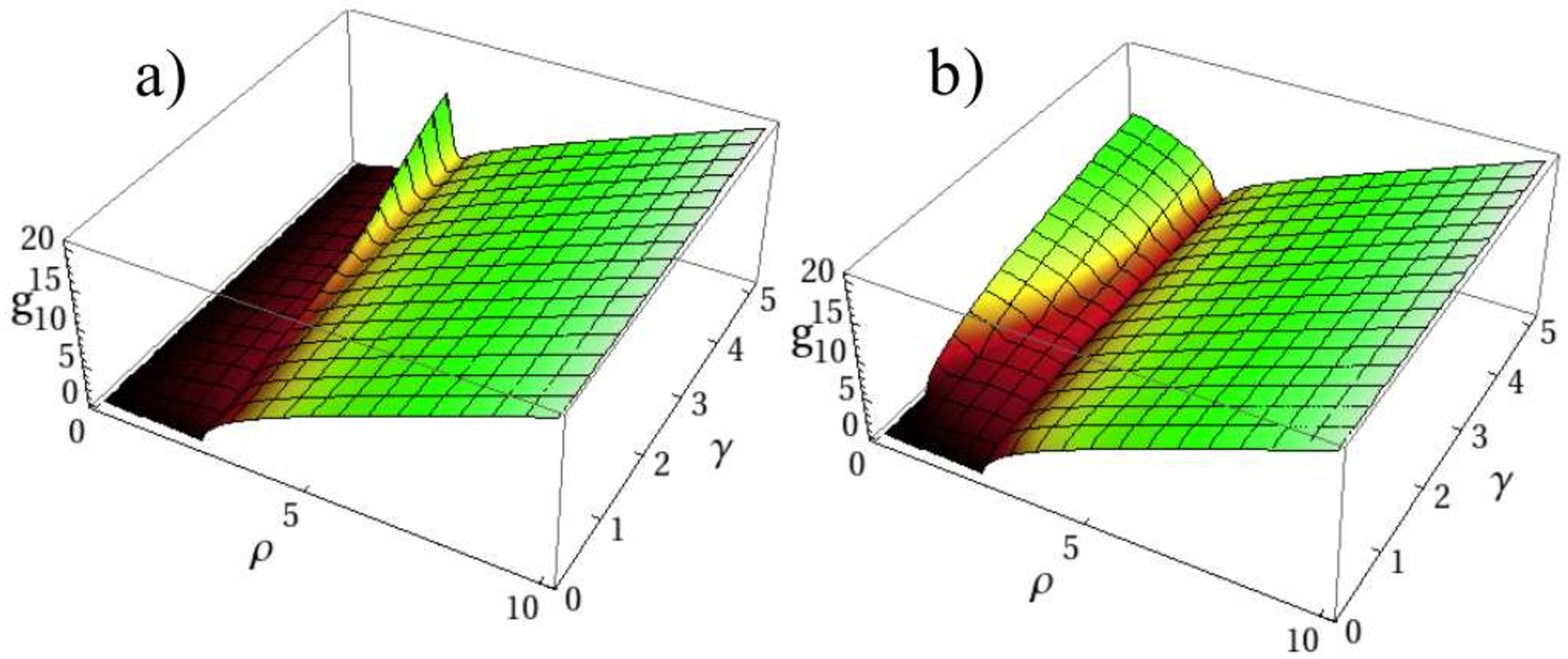}
\includegraphics[height=5.2cm]{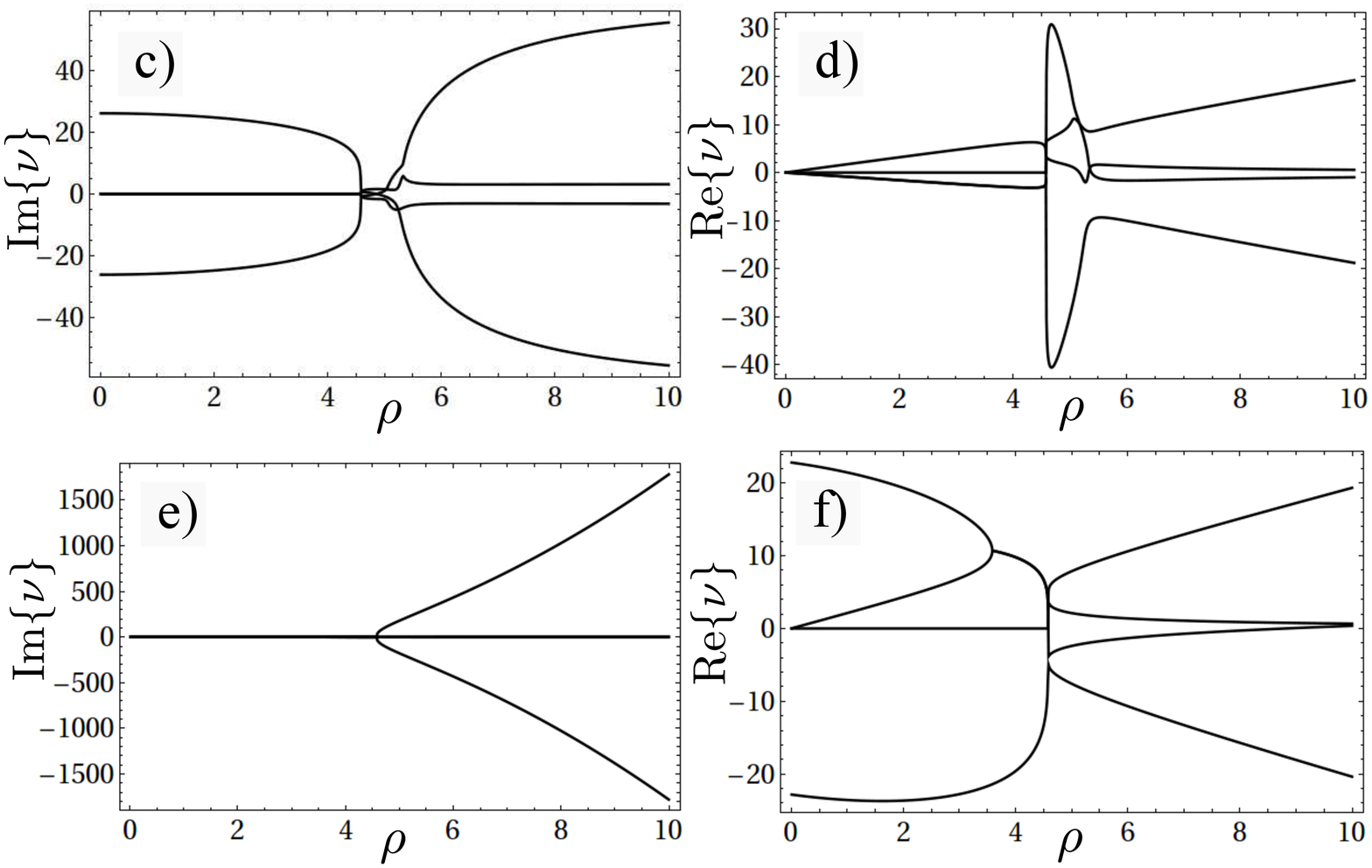}
\caption{(Color online) a) and b) show the instability gain
$g$ as function of $\rho>0$ and $\gamma$ for
$N=6$ for the $m^-$ and $m^+$ mode, respectively.
c), d), e) and f) show the
eigenvalues $\{\nu\}$ as function of $\rho$ ($0\leq\rho\leq 10$)
for $\gamma = 3$, and $N=21$.
c) and d) show the imaginary and real part of $\{\nu\}$ for the $m^-$ mode,
respectively, and e) and f) for the $m^+$ mode.}
\label{seis}
\end{figure}

Next, we split $\delta_{0}$ and $\delta_{1}$ into their real and
imaginary parts: $\delta_{0} = \alpha_{0} + i \beta_{0}$, $\delta_{1}
= \alpha_{1} + i \beta_{1}$.
We also decompose $a=x_{0}+i y_{0}$, $b=x_{1} + i y_{1}$.
After replacing into Eqs. (\ref{26}) and (\ref{27}) and after defining
$\vec{w}=(\alpha_{0}, \beta_{0}, \alpha_{1}, \beta_{1})^T$, we obtain an
equation of the form

\begin{figure}
\includegraphics[height=3.5cm]{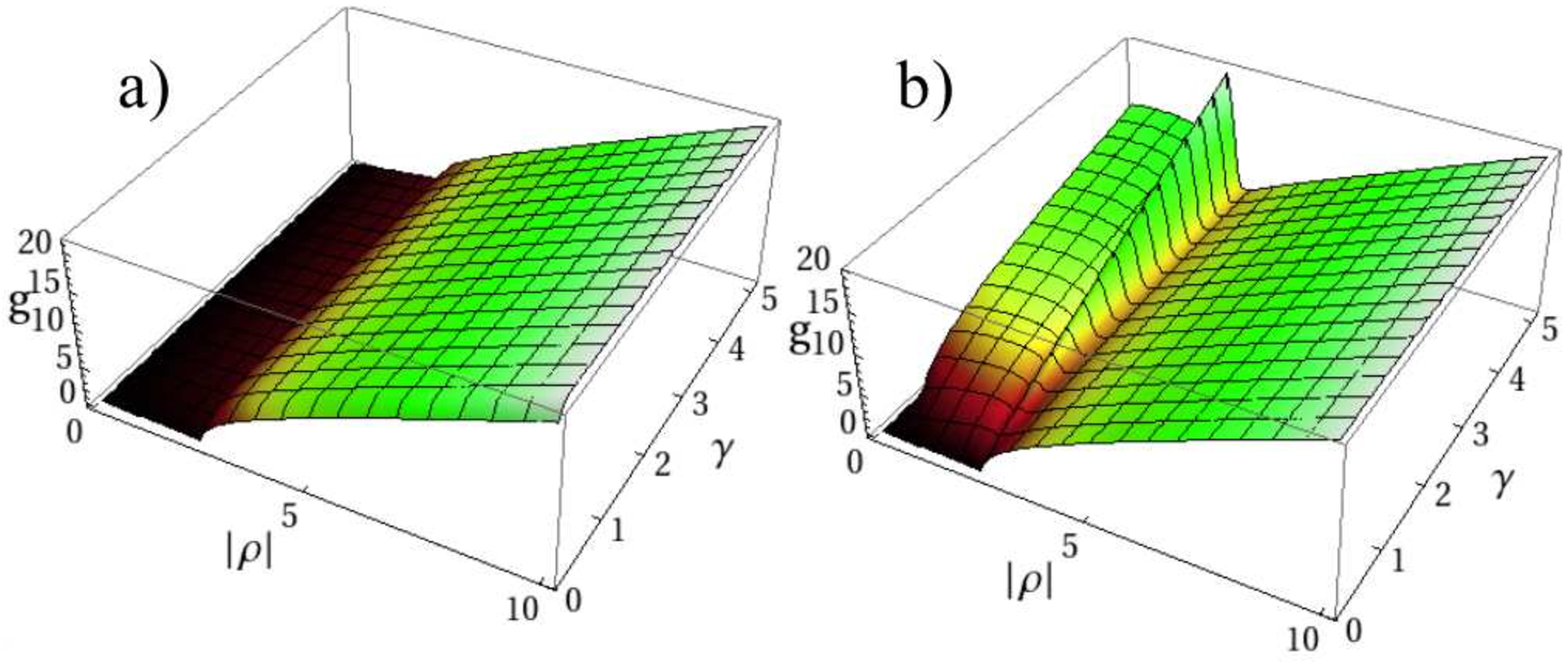}
\includegraphics[height=5.2cm]{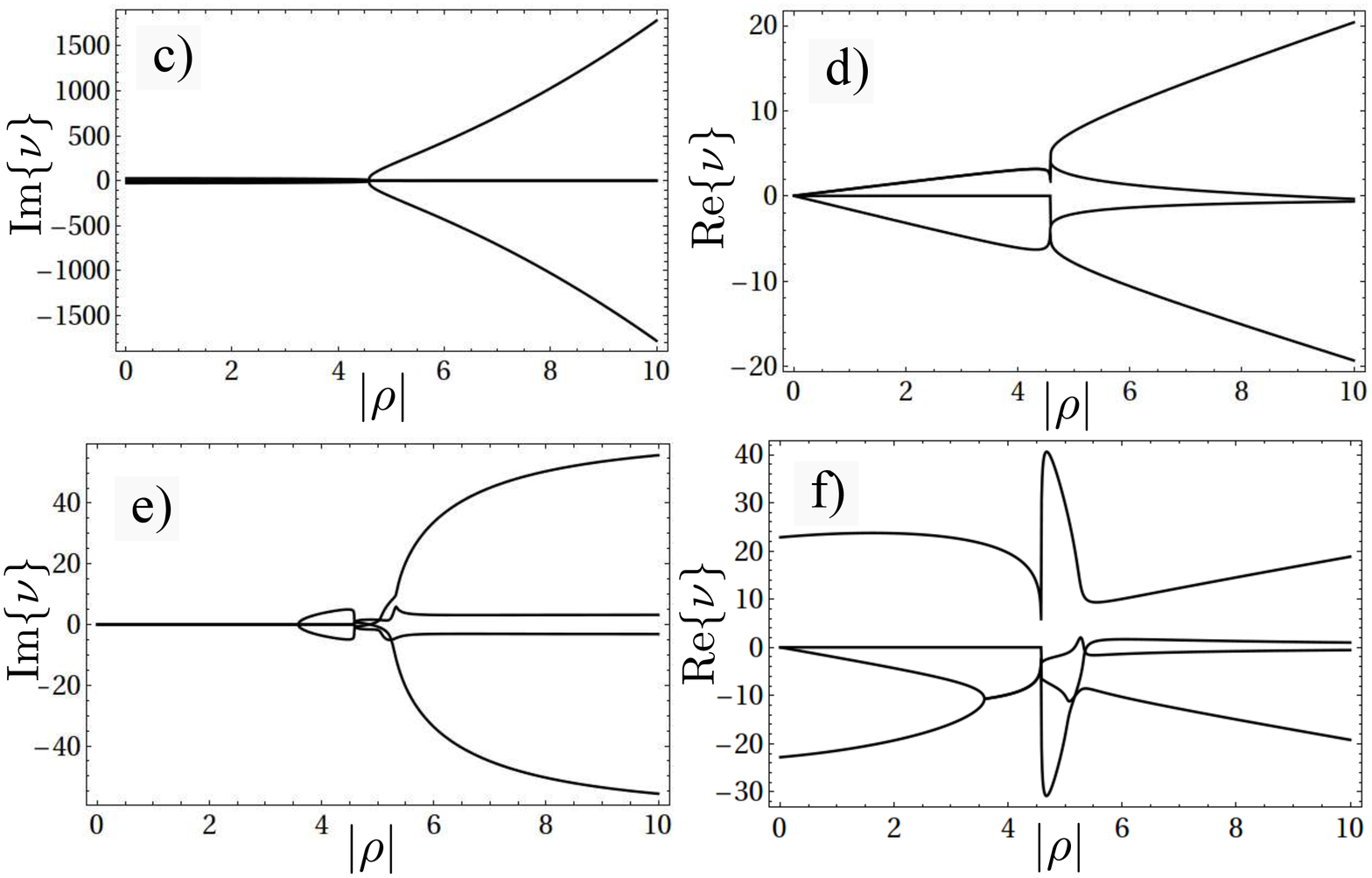}
\caption{(Color online) a) and b) show the
instability gain $g$ as function of $\rho<0$ and $\gamma$ for
$N=6$ for the $m^-$ and $m^+$ mode, respectively.
c), d), e) and f) show the eigenvalues $\{\nu\}$ vs. $|\rho|$ ($-10\leq\rho\leq 0$)
for $\gamma = 3$ and $N=21$. (c) and (d) show the imaginary and real part of $\{\nu\}$ for the mode $m^-$,
respectively, and (e) and (f) for the mode $m^+$ .}
\label{nueve}
\end{figure}

\be
\frac{d}{d z} \vec{w} = {\bf M}\ \vec{w},
\ee
where ${\bf M}=\{M_{i,j}\}$ is a $4 \times 4$ matrix with the components
$M_{1,3}=M_{2,4}=M_{3,1}=M_{4,2}=0$, and
\begin{eqnarray*}
M_{1,1}, \, M_{2,2} &=& - \rho \mp 2 \gamma x_{0} y_{0} \nonumber\\
M_{1,2}, \, M_{2,1}&=& \pm \lambda \mp \epsilon_{0}+\gamma (x_{0}^2-y_{0}^2) \mp
2 \gamma (x_{0}^2+y_{0}^2)\nonumber\\
M_{1,4}, \, M_{2,3} &=& \mp N C_{0}\nonumber\\
M_{3,2}, \, M_{4,1}&=& \mp C_{0}\nonumber\\
M_{3,3}, \, M_{4,4}&=&\rho \mp 2 \gamma x_{1}y_{1}\nonumber\\
M_{3,4}, \, M_{4,3}&=& \pm \lambda \mp \epsilon_{1}-2 C_{1}+\gamma (x_{1}^2-y_{1}^2) \mp 
2 \gamma (x_{1}^2+y_{1}^2) \nonumber\\
\end{eqnarray*}

The stability condition requires that the real part of all
eigenvalues $\{\nu\}$ of ${\bf M}$ be negative.
Thus, we define the instability gain $g$ as the real
part of the eigenvalue with the largest positive real part.

For the simple case with no gain and loss, it is possible to obtain the
eigenvalues in closed form:
\begin{eqnarray}
\nu_{1}^{+}&=&-4 N C_{0}^{2} + 2 \gamma (N+1) \sqrt{N} C_{0},\\
\nu_{2}^{+}&=& 0,
\end{eqnarray}
for the $m^+$ mode $\{\sqrt{N},1 \}$, and
\begin{eqnarray}
\nu_{1}^{-}&=&-4 N C_{0}^{2} - 2 \gamma (N+1) \sqrt{N} C_{0},\\
\nu_{2}^{-}&=&0,
\end{eqnarray}
for the $m^-$ mode $\{-\sqrt{N},1 \}$. We conclude that the $m^-$
mode is stable while the $m^+$ mode is stable provided
$\gamma < 2 \sqrt{N} C_{0}/(N+1)$. This defines a critical
nonlinearity, which depends on the size of the system, given by:
\be
\gamma_{c} = \frac{2 \sqrt{N} C_{0}}{(N+1)}.
\ee

However, numerical examination of the behavior
of the instability gain suggests that as soon as $\rho\neq 0$ the
nonlinear system becomes unstable.
For $\rho>0$, Figs. 6(a) and (b) show the
behavior of $g$ as function of $\rho$
and $\gamma$. For both modes, the interplay between nonlinearity and
gain and loss  causes destabilization of the modes. In the case of the
$m^+$ mode, this destabilization is bounded, at
least for $\rho<\rho_c$ and $\gamma<\gamma_c$. Out of this region, there
are bubble-like domains where $g$ increase abruptly. Nevertheless, in the
case of the $m^-$ mode, there are two mainly regions:
$\rho\lesssim\rho_c$, and $\rho\gtrsim \rho_c$. In the
former the mode is weekly unstable, otherwise, in the latter the mode
is highly unstable. Both regions are separated by a peak of instability.
Figs. 6(c), (d), (e), and (f) show the full outlook
associated to the distribution of eigenvalues of $M$ as function of $\rho$.
Moreover, Fig. 7 shows the same that Fig. 6, but considering $\rho<0$.

\begin{figure}
\includegraphics[height=7.cm]{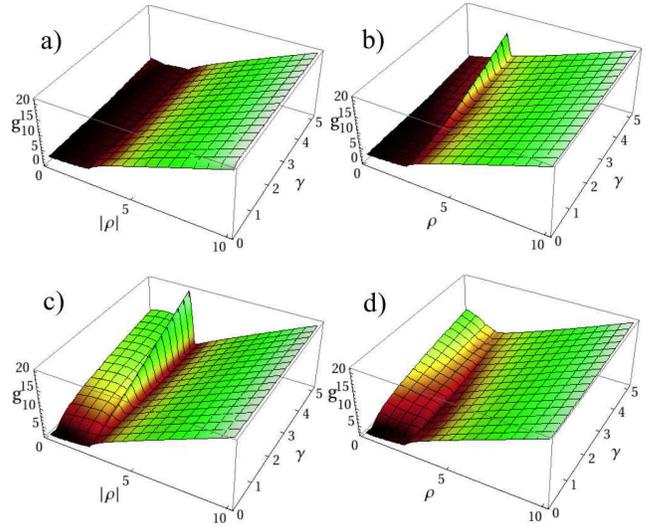}
\caption{(color online) Instability gain $g$ as function of $\rho$ and $\gamma$ for
$N=6$ using Eqs.~(31)-(34). (left) $\rho<0$ and (right) $\rho>0$. a)
and b) for the mode $m^-$. c) and d) for the mode $m^+$.}
\label{diez}
\end{figure}

The main result of this section is that, for a nonlinear dimer in a
stable regime, the addition of any amount of gain and loss will
destabilize the system. This feature will still hold for the general case,
as shown in the next section.

\subsection{Stability analysis in a general system}

Let us consider the stability problem for the general stationary modes
$\left\{A^s,\{B_n^s\}_{n=1}^N\right\}$ derived from Eqs. (\ref{eq:1}) and (\ref{eq:2}).  We introduce a linear perturbation of the form
$A(z) = (a^s+\delta_0(z))e^{i\lambda z}$, and
$B_n(z) = (b_n^s+\delta_n(z))e^{i\lambda z}$, with $a^s = x_0+iy_0$,
$b_n^s = x_n + iy_n$,
and
$\delta_n(z) = \alpha_n(z) + i\beta_n(z)$. Then, we obtain the following
linear system for the perturbation:
\begin{eqnarray}
\frac{d\alpha_0}{dz} &=& -(\rho_0+2\gamma x_0y_0)
\alpha_0
-C_0\sum_{j=1}^{N}\beta_j\label{34}\\&&
+[\lambda-\epsilon_0-\gamma(x_0^2+3y_0^2)]\beta_0,\nonumber\\
\frac{d\beta_0}{dz}&=&
(2\gamma x_0y_0-\rho_0)\beta_0+
C_0\sum_{j=1}^N\alpha_j \label{35}\\&&
+[\epsilon_0-\lambda + \gamma(3x_0^2 +
y_0^2)]\alpha_0, \nonumber\\
\frac{d\alpha_n}{dz} &=&
-(\rho_1+2\gamma x_ny_n)\alpha_n
-C_1(\beta_{n+1}+\beta_{n-1})\label{36}
\\&&
-C_0\beta_0
+[\lambda-\epsilon_1-\gamma(x_n^2+3y_n^2)]\beta_n,\nonumber\\
\frac{d\beta_n}{dz}&=&
(2\gamma x_ny_n-\rho_1)\beta_n
+C_1(\alpha_{n+1}+\alpha_{n-1})\label{37}
\\&&
C_0\alpha_0
+[\epsilon_1-\lambda + \gamma(3x_n^2 +y_n^2)]\alpha_n.\nonumber
\end{eqnarray}

Figure \ref{diez} shows the numerical results for the instability
gain obtained from this analysis.
We note that, the behavior of $g$ as function of $\rho$ and $\gamma$
is qualitatively the same that the one obtained using the analysis
for the reduced system. In fact, the instability bubbles are still
separated by a neighborhood around $\rho=\rho_c$. Further, for the
$m^-$ mode, we obtain that there is an small global increment of
the parameter $g$, which suggests higher levels of instability
respect to the reduced description. Nonetheless, numerically we found
that $m^-$ behaves ``stable'' in a bigger set of the
parameter space than $m^+$ does, since the stability region for $m^+$ is
bounded by $\rho_c$ and $\gamma_c$. This will be discussed in the
next section.

We can therefore conclude that a small addition of gain and loss
will destabilize the nonlinear multi-core array.


\section{Nonlinear dynamics and mode stabilization}

\begin{figure}
\includegraphics[height=2.2cm]{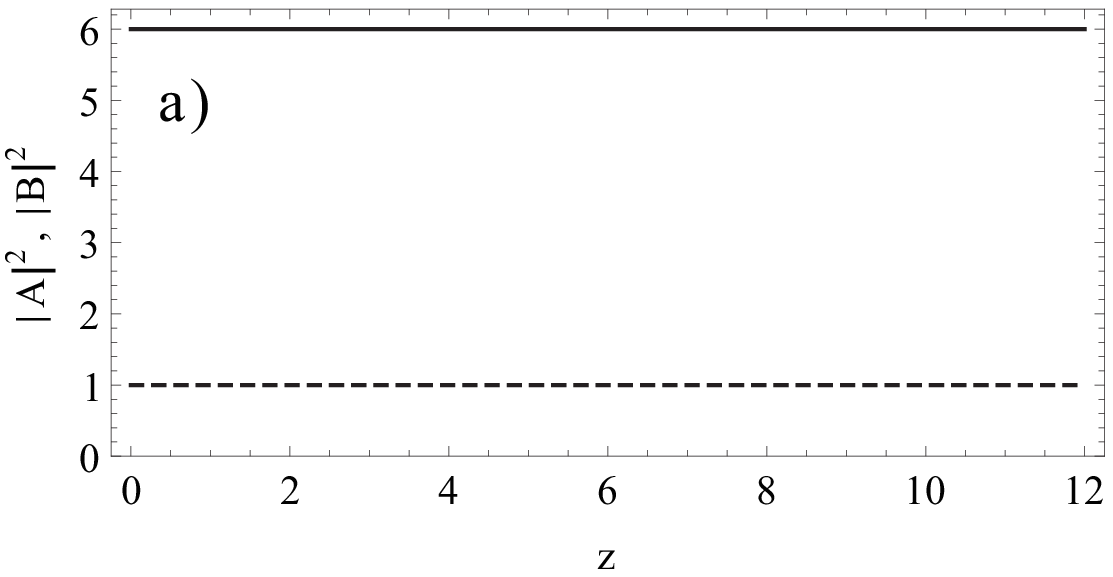}
\includegraphics[height=2.2cm]{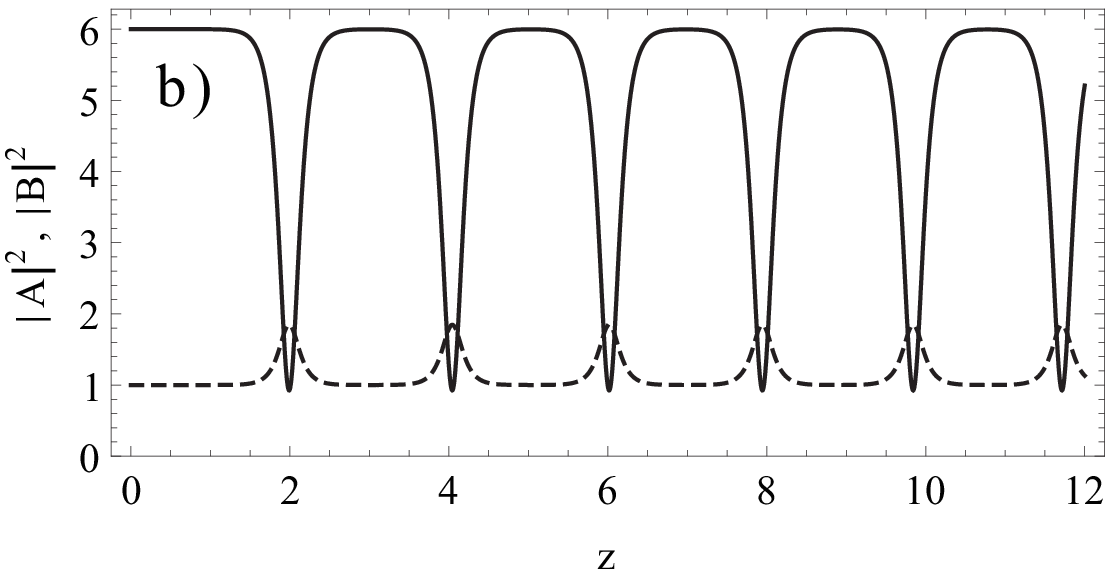}\\
\includegraphics[height=2.2cm]{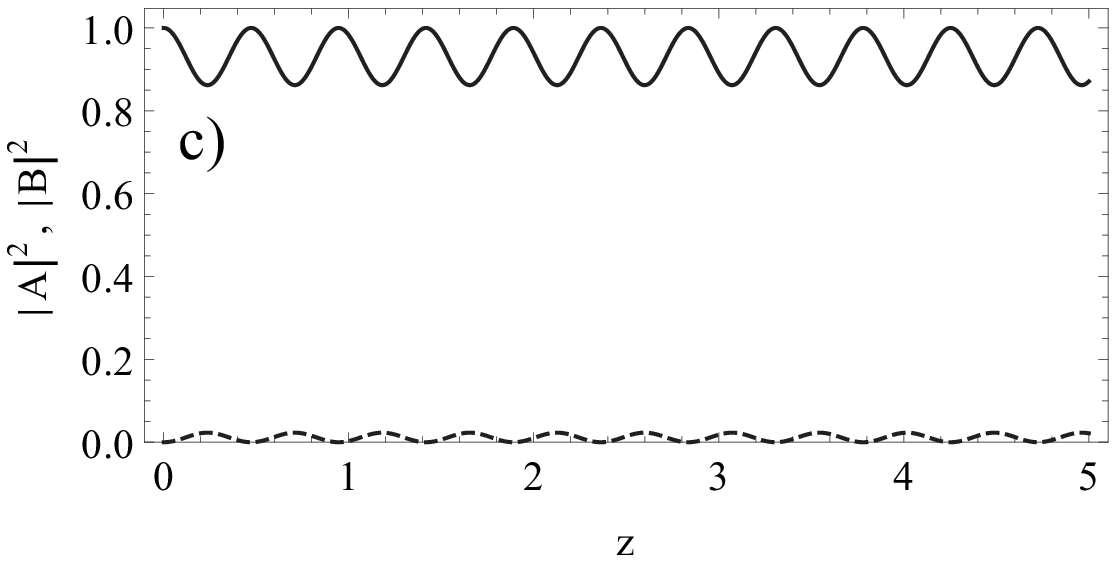}
\includegraphics[height=2.2cm]{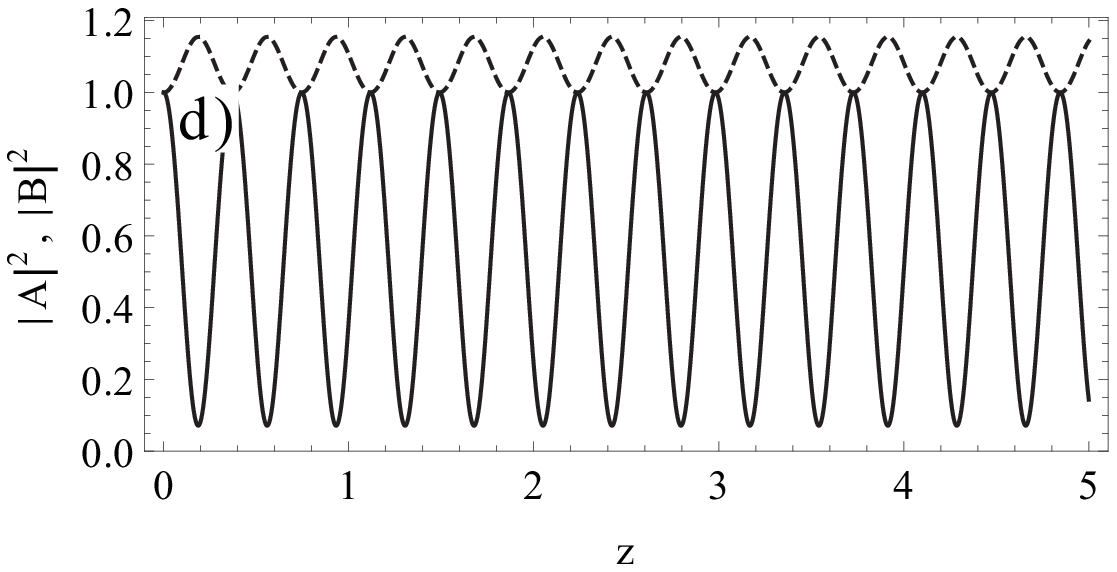}
\caption{Examples of numerical integration of Eqs. (1), (2)
in the nonlinear regime ($\gamma = 3$) for $N=6$ and $\rho=0$.
Continuous and dashed lines show $A(z)$ and $B_1(z)$,
respectively. Shown are: (a) $m^-$ mode, (b) $m^+$ mode,
(c) $A(0)=1$ and $B_j(0)=0$, (d) $A(0)=B_{j}(0) = 1$.}
\label{once}
\end{figure}

We now study numerically the system dynamics described by Eqs. (\ref{eq:1}), (\ref{eq:2}) in some interesting cases. First, we consider the  case $\rho=0$. Figure~\ref{once} shows the evolution for different initial conditions. Among these, we show the propagation of the nonlinear modes $m^\pm$. The $m^-$ mode displays an stable propagation along $z$, while the $m^+$ mode shows a modulation along $z$ as a consequence of an energy exchange between the core and the ring. On the other hand, when the initial excitation is either only at the core or in the whole array, the field oscillates periodically between the core and ring.

\begin{figure}
\includegraphics[height=2.2cm]{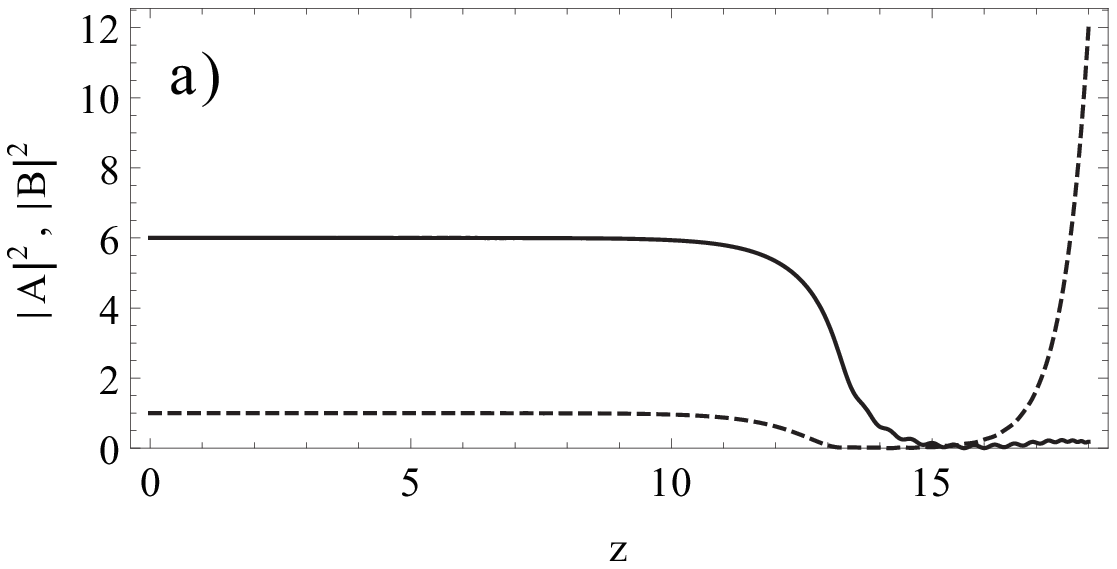}
\includegraphics[height=2.2cm]{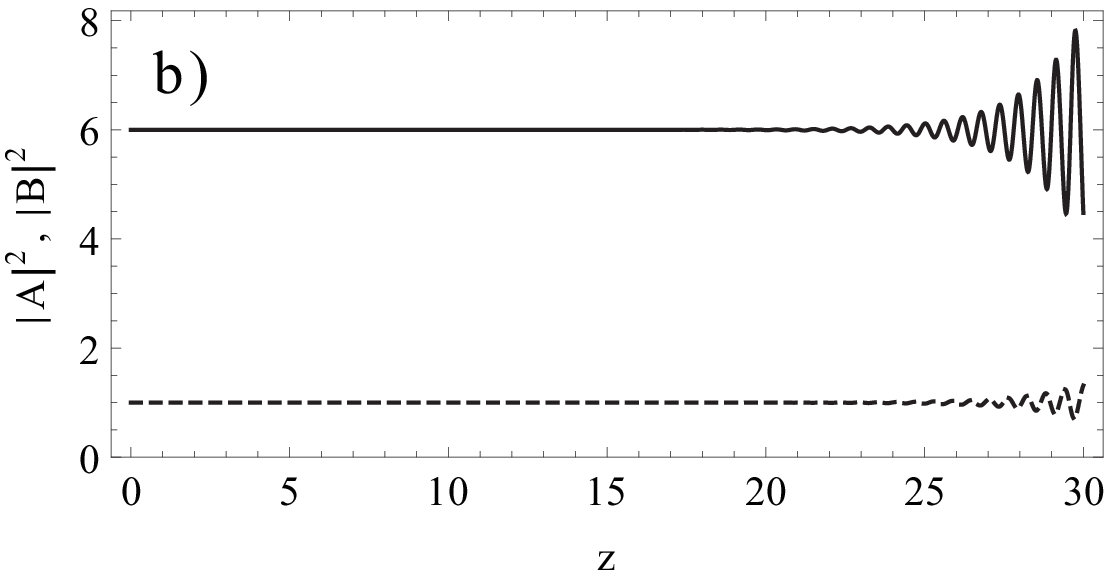}\\
\includegraphics[height=2.2cm]{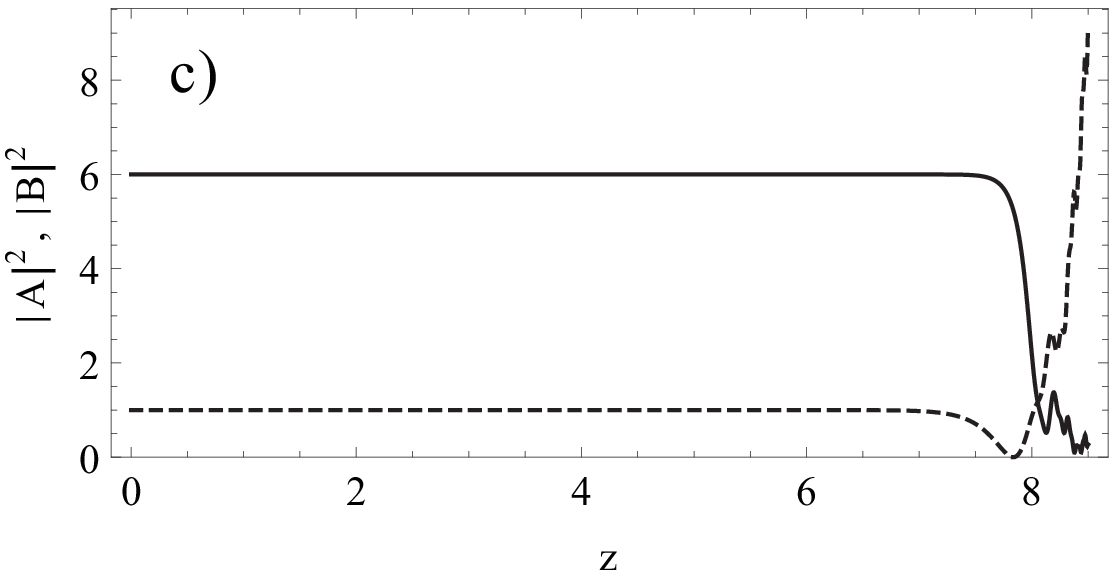}
\includegraphics[height=2.2cm]{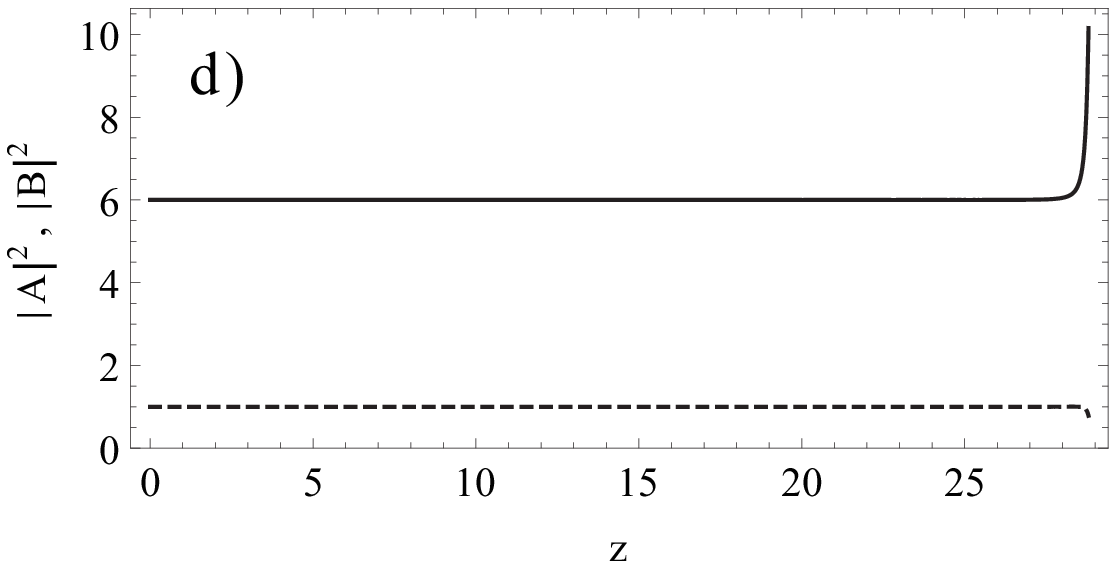}\\
\includegraphics[height=2.15cm]{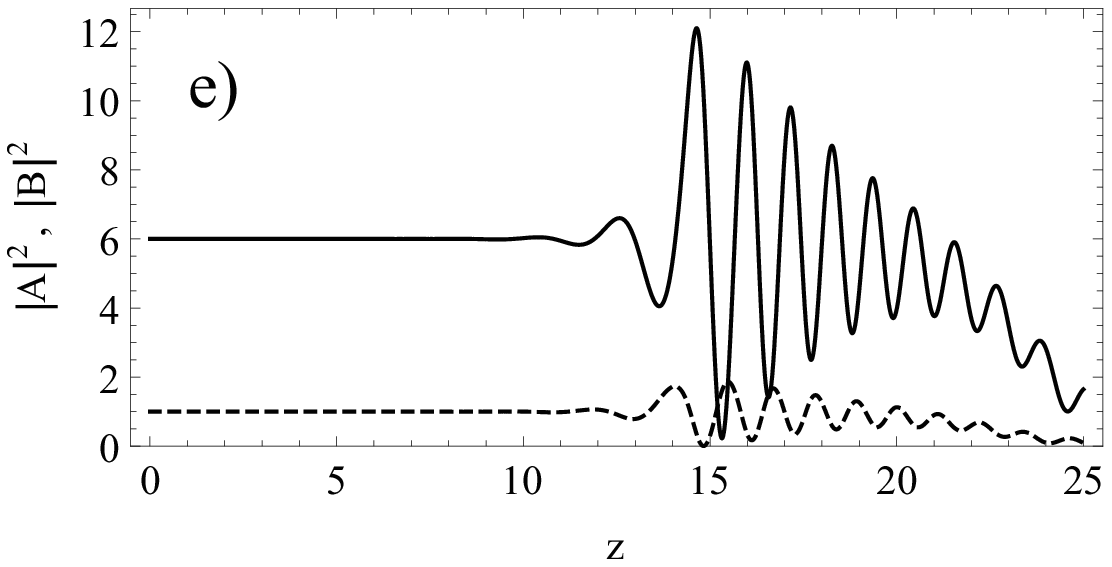}
\includegraphics[height=2.15cm]{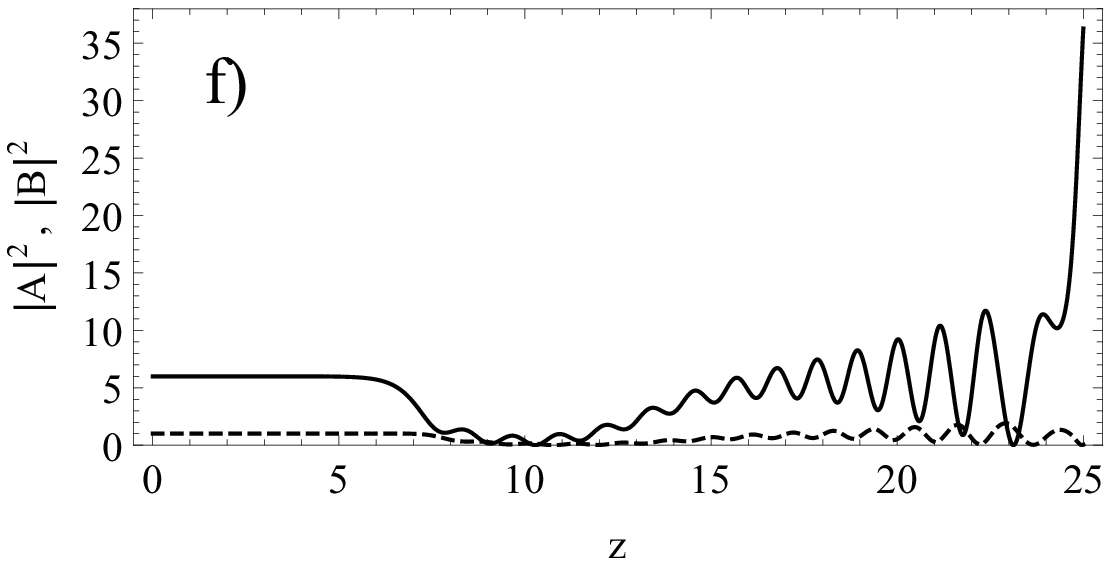}
\caption{Examples of numerical integration of Eqs.~(1), (2)
in the nonlinear regime for $N=6$ and $\rho\neq 0$.
Continuous and dashed lines show $A(z)$ and $B_1(z)$,
respectively. Left and right columns correspond to the cases
$\rho>0$ and $\rho<0$, respectively. Shown is  the mode $m^-$
in (a)-(d), and the mode $m^+$ in (e-f). Parameters are in (a, b)
$|\rho| = 1$ and $\gamma = 3$, (c, d) $|\rho|=\rho_c = \sqrt{6}$ and
$\gamma = 3$, and  (e, f) $|\rho| = 1$  and $\gamma=0.5<\gamma_c$.}
\label{doce}
\end{figure}

However, as soon as either gain or loss do not vanish, the evolution of the amplitudes $|A(z)|^2$ and $|B_{n}(z)|^2$ becomes unstable. In this case, the dynamics of almost any initial conditions shows an early divergence (for small $z$),
except for the  mode $m^-$, whose dynamics resembles a self-trapping state, up to the onset of instability. Figure~\ref{doce} shows some examples of the propagation for the $m^-$ mode for different parameters. Thus, simultaneous presence of nonlinearity, gain, and loss leads to destabilization of the system dynamics. Figure \ref{doce} shows that the optical field diverges sooner for $\rho>0$ than for $\rho <0$. This is a generic behavior of the
system for almost any initial condition, as shown below.

\begin{figure}
\includegraphics[height=5.cm]{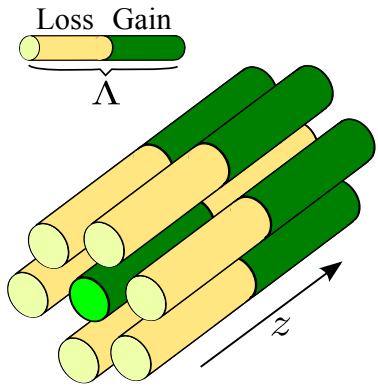}
\caption{(color online) Multi-core configuration with a square-like periodic
modulation of gain and loss along the propagation direction $z$.}
\label{trece}
\end{figure}

In order to understand a high level of instability for $\rho>0$, we employ the normal-mode analysis~\cite{flach}. First of all, the evolution equations for the optical field can be written in a general form as
\begin{equation}
i\frac{d\psi_n}{dz}+\sum_{m=1}^{N}V_{nm}\psi_{m}+\gamma|\psi_n|^2\psi_n=0,\label{35}
\end{equation}
where $V_{nm}$ represents the coupling between the waveguides $n$ and $m$
for $n\neq m$, while $V_{nn}$ characterizes local properties of the
$n$-th guide, such as local refractive index, gain, and loss.
Expanding $\psi_n(z)$ into normal modes (eigenfunctions for $\gamma = 0$) of Eq.~(35), such as
$\psi_n(z) = \sum_{m=1}^N\phi_{n,m}\Psi_{m}(z)$, with $\phi_{n,m}\in
\mathds{R}$, and using that $\left<\phi_m|\phi_l\right>=\sum_{n=1}^N\phi_{n,m}\phi_{n,l} =
\delta_{m,l}$, we can write Eq.~(\ref{35}) as
\begin{equation}
i\frac{d\Psi_{\nu}}{dz}+\lambda_{\nu}\Psi_{\nu}+\gamma\sum_{\nu_1
\nu_2\nu_3}I_{\nu\nu_1 \nu_2\nu_3}\Psi_{\nu_1}^*
\Psi_{\nu_2}\Psi_{\nu_3}=0,
\label{36}
\end{equation}
where $\lambda_{\nu}$ is the eigenvalue associated with the
eigenfunction $\psi_{\nu}$, and with the overlap integral
\begin{equation}
I_{\nu\nu_1\nu_2\nu_3} =
\sum_{n}\phi_{n,\nu}\phi_{n,\nu_1}\phi_{n,\nu_2}
\phi_{n,\nu_3}.
\end{equation}

Equation (\ref{36}) implies that nonlinearity induces an exchange of energy between
different linear modes. In our case, we have $N-1$ linear modes which have eigenvalue with imaginary part  $-\rho$.  Thus, even if a mode has a pure real propagation constant, for any finite value of $\gamma$, there will be
a finite contribution from $N-1$ unbounded modes in the dynamics. This, in addition to the effect of perturbations described  by Eqs.~(\ref{34})-(\ref{37}), will result in a high level of instability. This result is quite general, so
we conclude that it has to be generic for nonlinear systems with a complex eigenvalue spectrum.

\begin{figure}
\includegraphics[height=2.1cm]{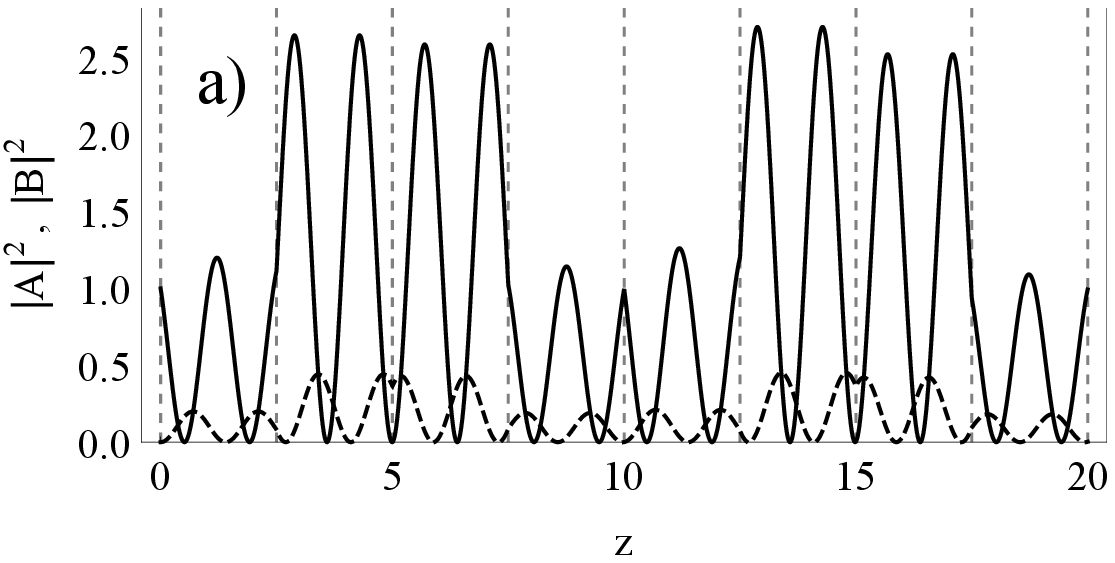}
\includegraphics[height=2.1cm]{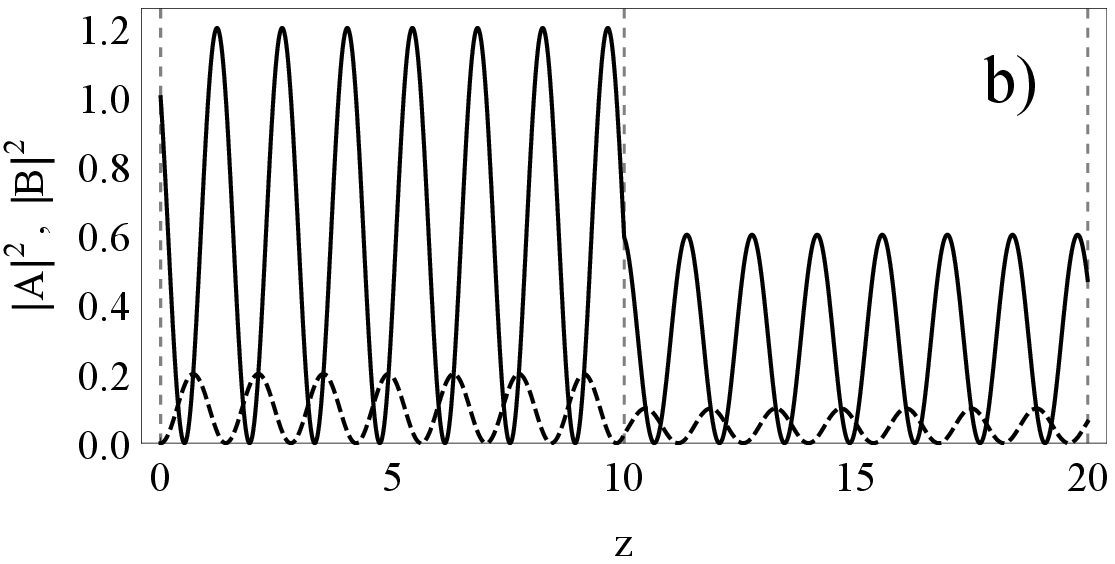}
\caption{Examples of numerical integration of Eqs.~(\ref{eq:1}), (\ref{eq:2})
in the linear regimen ($\gamma = 0$) for $N=6$ and $|\rho|=1$, with a square-like
modulation of $\rho$. Continuous and dashed lines mark $A(z)$ and $B_1(z)$,
respectively. Left and right columns correspond to the cases $\rho>0$ and $\rho<0$,
respectively. The initial conditions is $A(0)=1$ and $B_j(0)=0$ in both the cases.
(a) $\Lambda = 5$, (b) $\Lambda = 20$.}
\label{catorce}
\end{figure}

\begin{figure}
\includegraphics[height=2.1cm]{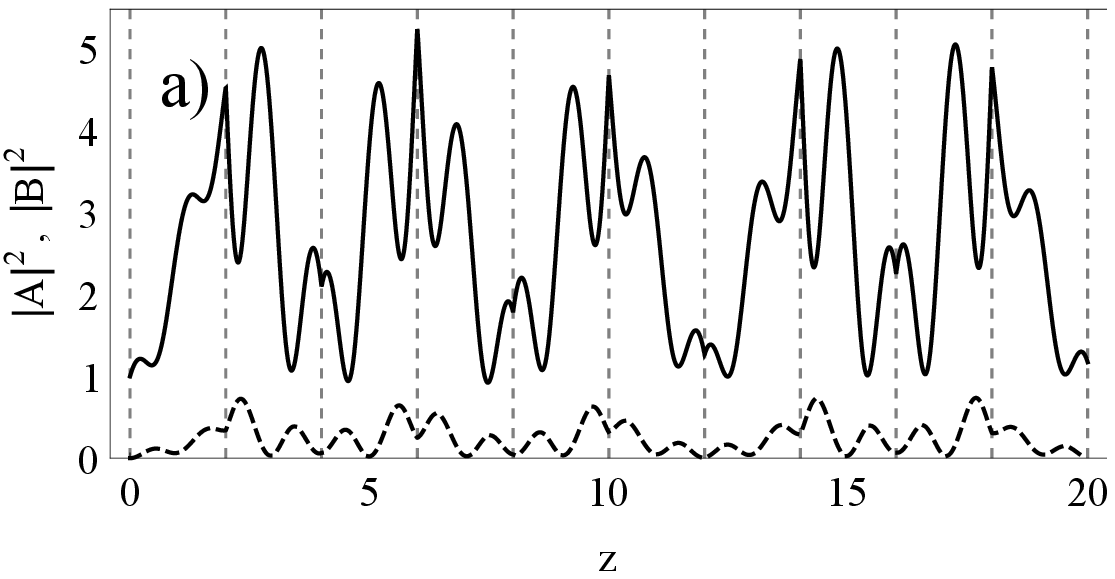}
\includegraphics[height=2.1cm]{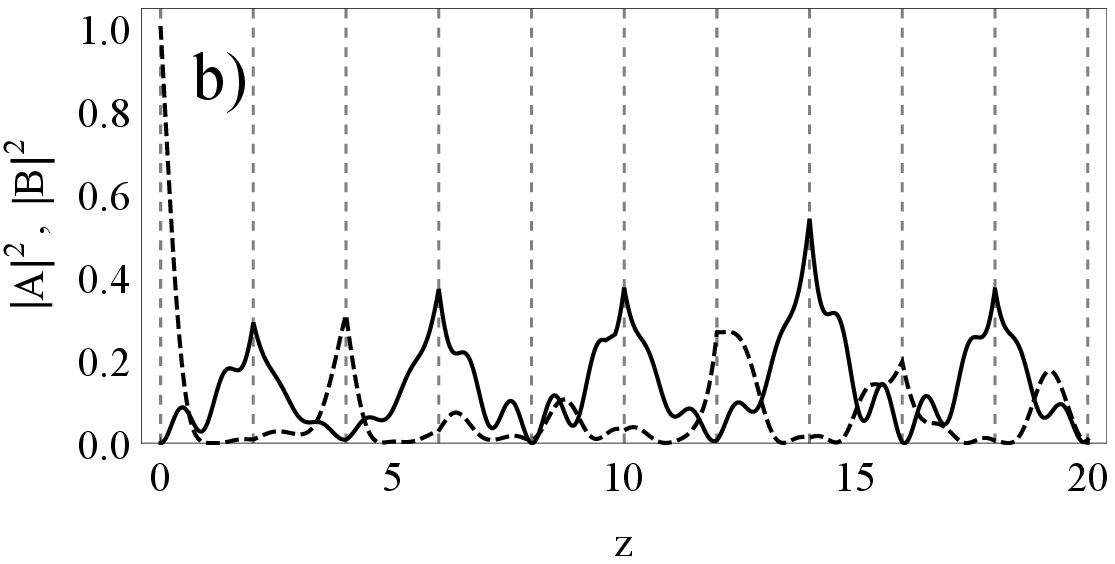}\\
\includegraphics[height=2.1cm]{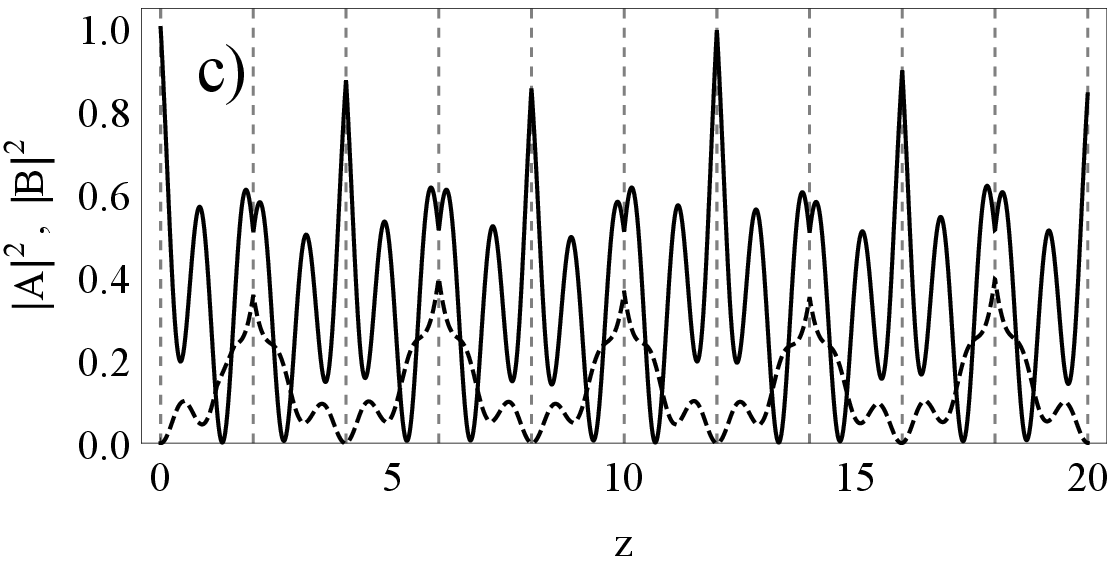}
\includegraphics[height=2.1cm]{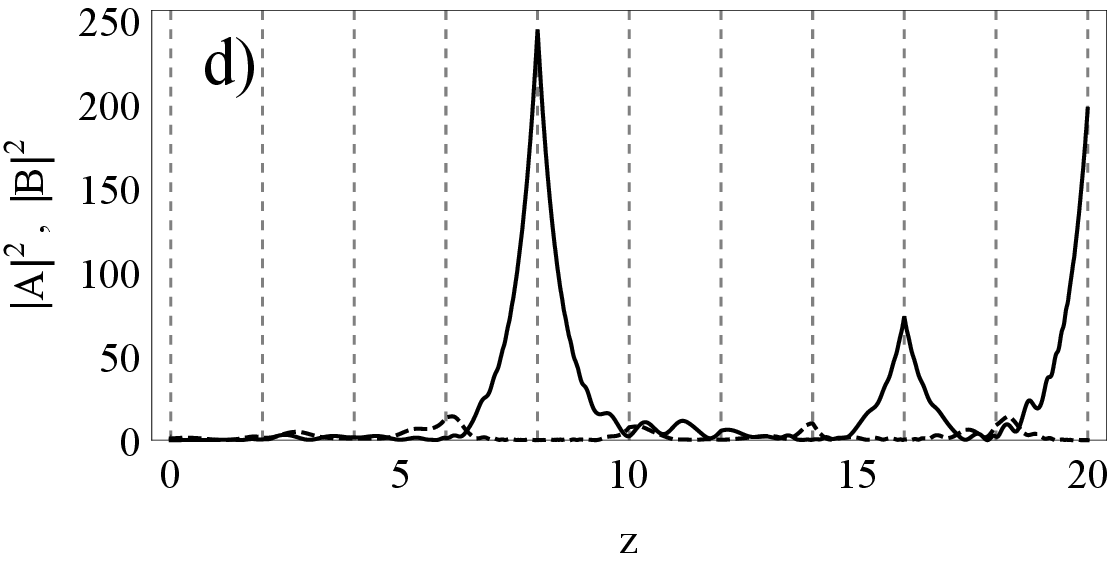}
\caption{Examples of numerical integration of Eqs.~(\ref{eq:1}), (\ref{eq:2})
in the nonlinear regime ($\gamma = 1$) for $N=6$ for $|\rho|=1$, considering a square-like
 modulation of $\rho$ with $\Lambda = 4$.
Continuous and dashed lines are associated to $A(z)$ and $B_1(z)$,
respectively. The parameters are: a) $\rho<0$, $A(0)=1$, $B_j(0)=0$,
b) $\rho<0$, $A(0)=0$, $B_1(0)=1$, $B_{j>1}(0)=0$,
c) $\rho>0$, $A(0)=1$, $B_j(0)=0$ and
d) $\rho>0$, $A(0)=1$, $B_1(0)=1$, $B_{j>1}(0)=0$.}
\label{quince}
\end{figure}

As we have seen, the nonlinear multi-core system becomes unstable in the presence of gain and loss. What this means is that energy does not dissipate and diffuse away at the same rate that it accumulates. To alleviate this problem we will introduce gain and loss terms whose sign will change periodically ~\cite{lee}. As a result, the spatial average of gain and loss  term will vanish, and the stable dynamics may be recovered. More specifically, we now take the gain and loss  parameter to be a function of $z$ (see Fig.~\ref{trece}). This kind of modulation has been implemented before by different authors, such as \cite{mod,localpt}, in the context of waveguide arrays, leading to an increased transport regime, and an unidirectional fractional phase exchange. We use here an square-like periodic modulation given by $\rho\rightarrow \rho(z) = \eta\, \text{square}(\Lambda z)$~\cite{mod}, where $\Lambda$ represents the period, and $\eta$ the intensity of the gain and loss. The divergences of optical fields depend on both gain and loss parameter $\rho$, and have the general form
$\sim e^{\text{sgn}\{\rho\}\upsilon z}$, where $\text{sgn}$ is the signum function, and $\upsilon$ is a measure of the strength of the divergence. Thus, when the sign of $\rho$ along $z$ changes, the dynamics goes from an exponential increment to an exponential decrement, or viceversa.

Examples of the effects of this modulation on the dynamics are shown in Fig.~\ref{catorce} and Fig.~\ref{quince} for both linear and nonlinear regimes, respectively. In the linear regime, the propagation constants are not modified, and we only
observe effects in the amplitude of the waves. How amplitude will respond depends on the ratio between the propagation constant $\lambda$ and period $\Lambda$ of the modulation. On the other hand, in the nonlinear regime we found that the modulation induces an effective dynamic stabilization, where the field remains bounded along the propagation direction. It is interesting to notice that this effect is also present for initial conditions that do not have any particular
symmetry, such as the case of Figs.~\ref{quince} (b) and (d).

\section{Conclusions}

We have studied the dynamics of nonlinear multi-core waveguiding structures with balanced gain and loss. For the linear regime, we have shown that the bounded dynamics can be observed in the
limit of an effective waveguide dimer and when gain is placed in the
core of the multi-core structure. Thus, the dynamics can be reduced to that  of an effective ${\cal PT}$-symmetric waveguide dimer with an asymmetric coupling. Within this reduction, we have computed the eigenvalues and eigenvectors of the structure and found the critical value of gain and loss for an onset of the ${\cal PT}$-symmetry-breaking instability.

In the nonlinear case, we have found and analyzed nine stationary modes, four of which bifurcate from the linear modes.
For these modes, we have found that the propagation constant remains the same as for the linear case but it gets shifted, while the eigenvectors are found to have the same envelope as in the linear case, except by a constant factor which depends on the power and the number of waveguides in the system. For these modes, we have conducted the stability analysis, and found that the modes are all unstable in the presence of nonlinearity, gain, and loss. We have found a critical
parameter for balanced gain and loss separating the regions of low and high instability. Furthermore, we have revealed that the  stabilization of nonlinear modes can be achieved by applying a spatially periodic modulation of gain and loss, and we have examined the corresponding bounded dynamics for all initial conditions. We believe that an experimental realization of these findings might help resolve issues related to optical energy transport through multi-core waveguiding structures.

\section*{Acknowledgements}

This work was partially supported by FONDECYT grant 1120123, Programa
ICM P10-030-F, Programa de Financiamiento Basal de CONICYT
(FB0824/2008), and Australian Research Council.  
A.J.M. acknowledges partial support from CONICYT (BCH72130485/2013)

\end{document}